\documentclass[11pt]{article}
 
\usepackage{bm}
\usepackage[
backend=biber,
style=apa,
natbib=true,
]{biblatex}
\usepackage{booktabs}
\usepackage{mathtools} % Better math support
\usepackage{amssymb} % Fancy math symbols
\usepackage{xcolor}
\usepackage[margin=1in]{geometry}
\usepackage{microtype} % Make the compiled document prettier
\usepackage{lineno} % Add line numbering
\usepackage{subcaption} % Add subfigures
\usepackage{graphicx}
\usepackage{hyperref} % Turn references, etc into clickable links
\usepackage{algorithm}
\usepackage{algpseudocode}

\hypersetup{%
  colorlinks = true, % false: boxed links; true: coloured links
  citecolor = blue,
  urlcolor = blue,
  bookmarks = true % Adds a navigation menu in the pdf reader
}

\newcommand{\inst}[1]{\vspace{1pt} \unskip\(^{#1}\)}

\begin{document}

\title{\vspace{-2cm} Quantile based modelling of diurnal temperature range with the
  five-parameter lambda distribution}

\author{Silius M. Vandeskog\inst{1} \and
  Thordis L. Thorarinsdottir\inst{2} \and
  Ingelin Steinsland\inst{1} \and
  Finn Lindgren\inst{3}}

\date{}

\maketitle

\noindent \inst{1} Norwegian University of Science and Technology (NTNU) \\
\inst{2} Norwegian Computing Centre \\
\inst{3} The University of Edinburgh

\begin{abstract}
  Diurnal temperature range is an important variable in climate science that can provide
  information regarding climate variability and climate change. Changes in diurnal temperature
  range can have implications for hydrology, human health and ecology, among others.
  Yet, the statistical literature on modelling diurnal temperature range is lacking.
  In this paper we propose to model the distribution of diurnal temperature range using the
  five-parameter lambda (FPL) distribution.
  Additionally, in order to model diurnal temperature range with
  explanatory variables, we 
  propose a distributional quantile regression model that combines quantile regression with marginal
  modelling using the FPL distribution.
  Inference is performed using the method of quantiles.
  The models are fitted to 30 years of daily observations of diurnal temperature range from 112
  weather stations in the southern part of Norway. The flexible FPL distribution shows great promise
  as a model for diurnal temperature range, and performs well against competing models.
  The distributional quantile regression model is fitted to diurnal temperature range data using geographic,
  orographic and climatological explanatory variables.
  It performs well and captures much of the spatial variation in the distribution of diurnal
  temperature range in Norway.
\end{abstract}

{\bf Keywords:} Diurnal temperature range, five-parameter lambda distribution, method of quantiles,
  distributional quantile regression

\section{Introduction}
\label{sec:introduction}

In this paper we develop distributional models for diurnal temperature range, which is the
difference between daily maximum and minimum temperature. The fourth IPCC assessment report
identified diurnal temperature range as a key uncertainty factor \citep{ipcc2007science}, and the fifth
report described a substantial knowledge gap surrounding this climate variable
\citep{ipcc2013science}. While more effort has since been made to understand diurnal temperature
range, the literature on it is still lacking \citep{thorne16_reass_chang_diurn_temper_range,
  thorne16_reass_chang_diurn_temper_range2, ipcc2021, sun19_global_diurn_temper_range_dtr}.
Diurnal temperature range can be used
as an index of radiative forced climate change, and as a useful index for assessing the output of
general circulation models \citep{braganza04_diurn_temper_range_as_index}. Additionally, it has
been shown that diurnal temperature range is linked to human health conditions such as the risk of
influenza \citep{park20_effec_temper_humid_diurn_temper}, risk of stroke
\citep{vered20_high_ambien_temper_summer_risk}, and overall health and mortality
\citep{cheng14_impac_diurn_temper_range_human_healt,lim15_diurn_temper_range_short_term}.
Changes in diurnal temperature range can also be of large importance within ecology
\citep{peng13_asymm_effec_daytim_night_time, vasseur14_increas_temper_variat_poses_great,
  Kovi&2016, Henry2007} and hydrology \citep{hanssen2016}, among others.
Despite these areas of usage, to the best of our knowledge, there do not exist any attempts
at statistical modelling of the distribution of diurnal temperature range in the literature.
In their recent work of reassessing changes in diurnal temperature range worldwide,
\citet{thorne16_reass_chang_diurn_temper_range} conclude that there is ``only
\textit{medium confidence} in the magnitude of reductions in diurnal temperature range since
1950'' and that ``there is \textit{low confidence} in trends and multidecadal variability in
diurnal temperature range prior to 1950''. Thus, more knowledge about diurnal temperature range is
needed.

In this paper we propose to model the marginal distribution of diurnal temperature range with
the five-parameter lambda (FPL) distribution \citep{gilchrist2000}. The FPL distribution is an extension of the
four-parameter generalised lambda distribution \citep{ramberg1974gld}, itself an extension of
Tukey's three-parameter lambda distribution \citep{tukey1962}. This family of distributions has seen
infrequent use within the statistical literature. Some areas of usage for the FPL distribution have
been income modelling \citep{tarsitano2004income} and reliability analysis
\citep{nair2013,ahmadabadi2012process}. The FPL distribution is tightly linked to the
generalised Pareto distribution \citep[e.g.][]{coles2001introduction}, as its quantile
function is equal to the difference between two generalised Pareto quantile functions (see Section~\ref{sec:fpld}). The
generalised Pareto distribution is often used for estimating extremely large or small quantiles,
and has been much used for modelling both the upper and lower tails of temperature distributions
\citep[e.g.][]{stein21_param_model_distr_with_flexib,rohrbeck21_spatio_tempor_model_red_sea}.
Thus, the FPL distribution is a natural choice for 
modelling the difference between daily maximum and daily minimum temperature if one considers
these as extreme upper and lower quantiles in the daily temperature distribution.

The FPL distribution can be used for modelling the distribution of diurnal temperature
range in locations with available observations. However,
most locations do not contain any available temperature data. Thus, it is also of interest
to model diurnal temperature range in locations without daily temperature observations, using a regression
model. Most classical regression models focus on estimating the
conditional mean of a distribution, given a set of explanatory variables. However, the distribution of
diurnal temperature range is complex, and its variance, skewness and kurtosis vary in
space (see Section~\ref{sec:data}). Thus, a regression model for the mean would 
not provide enough information about diurnal temperature range to be of
much use. An alternative to regression on the mean is quantile regression \citep{koenker2005quantileRegression},
where one models a set of conditional quantiles given some explanatory variables. Quantile
regression is based on fewer assumptions than mean regression, and it allows for more flexible
modelling of complex distributions. However, it can often lead to quantile crossing, meaning that certain
combinations of the explanatory variables lead to non-monotonic quantile functions
\citep{rodrigues2017regression, cannon2011quantileReg,
  bondell10_noncr_quant_regres_curve}. Furthermore, quantile regression can only estimate a finite
set of quantiles, and it does not lend an easy way of estimating distributional properties like moments. 
An alternative that has gained more popularity in recent years is distributional regression, where
one attempts to model the entire conditional distribution given a set of explanatory variables
\citep[e.g.][]{klein15_bayes_struc_addit_distr_regres, henzi20_isoton_distr_regres,
  schlosser19_distr_regres_fores_probab_precip}. As stated by \citet{hothorn14_condit_trans_model},
this should be the ultimate goal of any regression analysis. However, most distributional regression
models can be somewhat complex and computationally demanding.

Here, we propose a conceptually simple and highly parallelisable
distributional regression model, based on a combination of quantile regression and marginal modelling
with the FPL distribution. Parameter estimation for the FPL distribution is performed using the
method of quantiles \citep[e.g.][]{koenker2005quantileRegression}, which is based on minimising
the distance between a quantile function and a set of estimated quantiles. A thorough
description of this estimation method is presented, and it is compared to competing estimation methods for the
FPL distribution, both in a simulation study, and using real temperature observations.
In order to ease parameter interpretation and numerical inference, a novel reparametrisation of
the FPL distribution is developed.
The marginal FPL model and the distributional quantile regression model are fitted to
thirty years of daily temperature observations from 112 weather stations in
southern Norway, including both coastal and inland stations over a large range of altitudes. 
In order to properly evaluate the model fits, a closed-form expression for the
continuous ranked probability score \citep[CRPS,][]{matheson76_scorin_rules_contin_probab_distr}
with an FPL forecast distribution is developed.
This new modelling framework provides a rigorous alternative to analyse diurnal temperature range
and its observed variation in time and space compared to current empirical approaches
\citep[e.g.][]{wang17_asymm_urban_daily_air_temper_cycle,
  vinnarasi17_unrav_diurn_asymm_surfac_temper, shelton21_long_term_trend_diurn_temper,
  sun21_chang_diurn_temper_range_over}.

The remainder of the paper is organised as follows. Section~\ref{sec:data} introduces daily
temperature data from the southern part of Norway, and associated explanatory
variables. Section~\ref{sec:models} provides 
a motivation for the choice of modelling diurnal temperature range with the FPL distribution, and presents some
of the 
properties of the distribution. The distributional quantile regression model is also developed here. In
Section~\ref{sec:inference}, the method of quantiles
and two other competing inference methods are described, and a closed-form expression for the CRPS
with
an FPL forecast distribution is developed.
In Section~\ref{sec:simulation-study}, a
simulation study is performed, where we compare the method of quantiles with two competing inference
methods. Finally, we apply our models to Norwegian diurnal
temperature range data and evaluate the model fits in Section~\ref{sec:results}. The paper concludes with a short discussion in
Section~\ref{sec:discussion}.

\section{Data}
\label{sec:data}

\begin{figure}
  \centerline{\includegraphics[width=.95\linewidth]{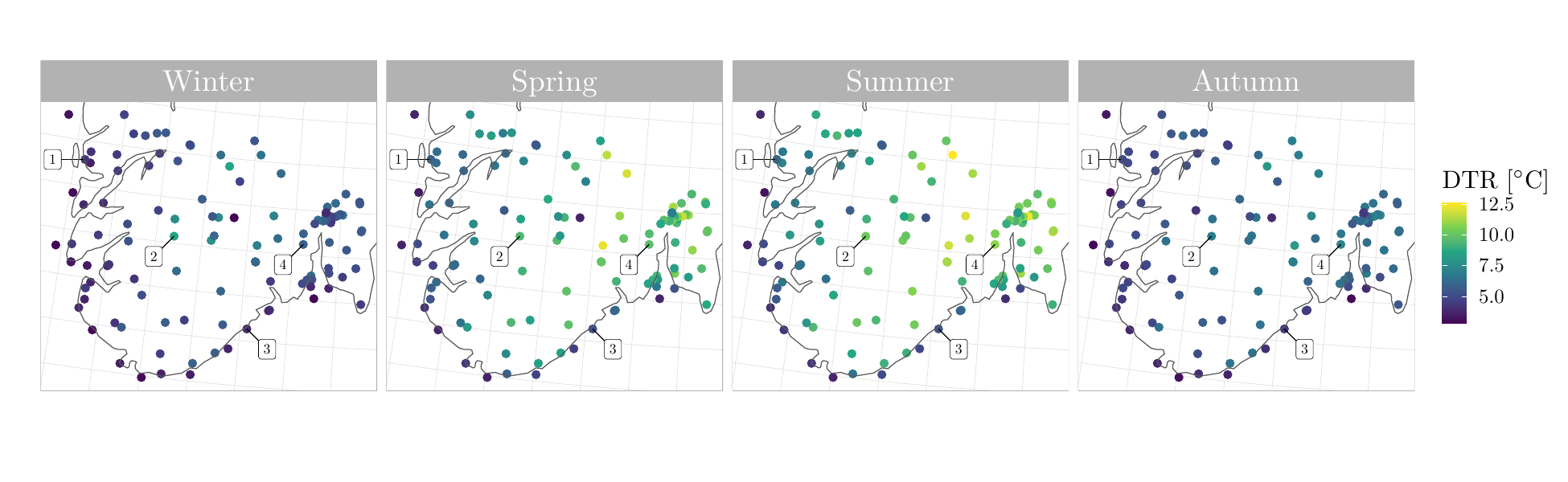}}
  \caption{Empirical season medians of diurnal temperature range (DTR) at the 112
    weather stations in our data set. The locations of the four stations from
    Figures~\ref{fig:range-histograms}, \ref{fig:local_distribution_fits} and
    \ref{fig:regional-out-of-sample} are presented in all the plots. These
    are: (1) Flesland, (2) Hovden - Lundane, (3) Lyngør fyr and (4) Sande - Galleberg.}
  \label{fig:median_values}
\end{figure}

The analysis in this paper is based on daily time series of air temperature observations from a
set of weather stations in southern Norway. The data are openly available from \citet{eklima}.
For each weather station, daily minimum and maximum
temperatures between 18-18 UTC are used to find time series of diurnal
temperature range. Data is downloaded from the thirty year time period from 1 January
1989 to 31 December 2018. Two thirds of the weather stations were already established in 1989, and
the ages of the remaining stations are almost uniformly distributed between one and thirty years.
Some stations are too recently established to be useful for our purposes, and others contain large
amounts of missing data. Data cleaning is therefore
performed by removing all weather stations that contains less than 180 observations from any of the
four seasons of the year (winter:
December-February; spring: March-May; summer: June-August; autumn: September-November). By
cleaning the data we reduce the number of weather stations from an initial 133 
to a new value of 112 stations. The locations of these are displayed in
Figure~\ref{fig:median_values} together with median diurnal temperature range for each season.
These 112 stations span altitudes from 0 to 1900 meters above sea level, and contains observations
from tundra, subarctic and oceanic climates
\citep{kottek06_world_map_koepp_geiger_climat_class_updat}.

Our modelling framework does not specifically account for measurement and
round-off errors. Obvious errors, resulting in negative diurnal
temperature range, are removed from the data. Due to the existence of
negative data, it is expected that there also are erroneous data
among the positive range values. However, accounting for these errors is outside the scope of this paper.

Six explanatory variables are used for modelling diurnal temperature range: easting, northing,
distance to the open sea and altitude, in addition to the historical
mean and variance of daily mean temperature at each location. This is estimated using the records of
daily temperature observations from each station. Exploratory analysis finds evidence that the
marginal distribution of diurnal temperature range can be approximated as being constant within each
season (results not shown), but that it varies between the seasons. 
Historical mean and variance of the
daily mean temperature are therefore computed for each season separately. The
distance to the open sea is derived from a digital elevation model of
Norway, with resolution \(50 \times 50\) m\(^2\), published by the Norwegian Mapping Authority
(\url{https://hoydedata.no}). Time series of daily mean
temperature, along with longitude, latitude and altitude are freely
available from \citet{eklima}. Easting and northing are based on UTM 32 coordinates.

\begin{figure}
  \centerline{\includegraphics[width=.9\linewidth]{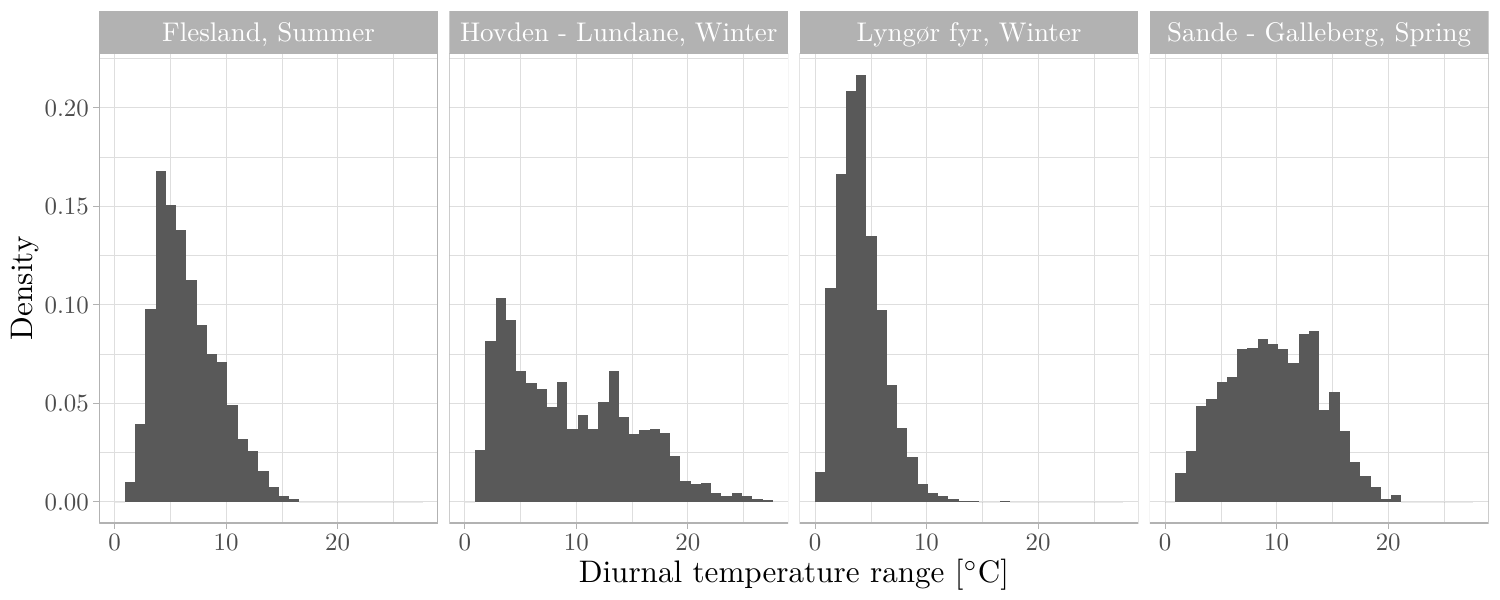}}
  \caption{Histograms of diurnal temperature range for four selected
    weather stations and seasons.}
  \label{fig:range-histograms}
\end{figure}

Figure~\ref{fig:median_values} shows the median of diurnal temperature
range for each season and each weather station.
It reveals clear seasonal and spatial patterns. In particular, the values appear higher
during spring and summer than during winter and autumn. Similar patterns are also found when examining
other quantiles. Histograms of diurnal temperature range for four selected stations and seasons are
presented in Figure~\ref{fig:range-histograms}. We observe
considerable differences in the shapes of the histograms.

\begin{figure}
  \centerline{\includegraphics[width=.8\linewidth]{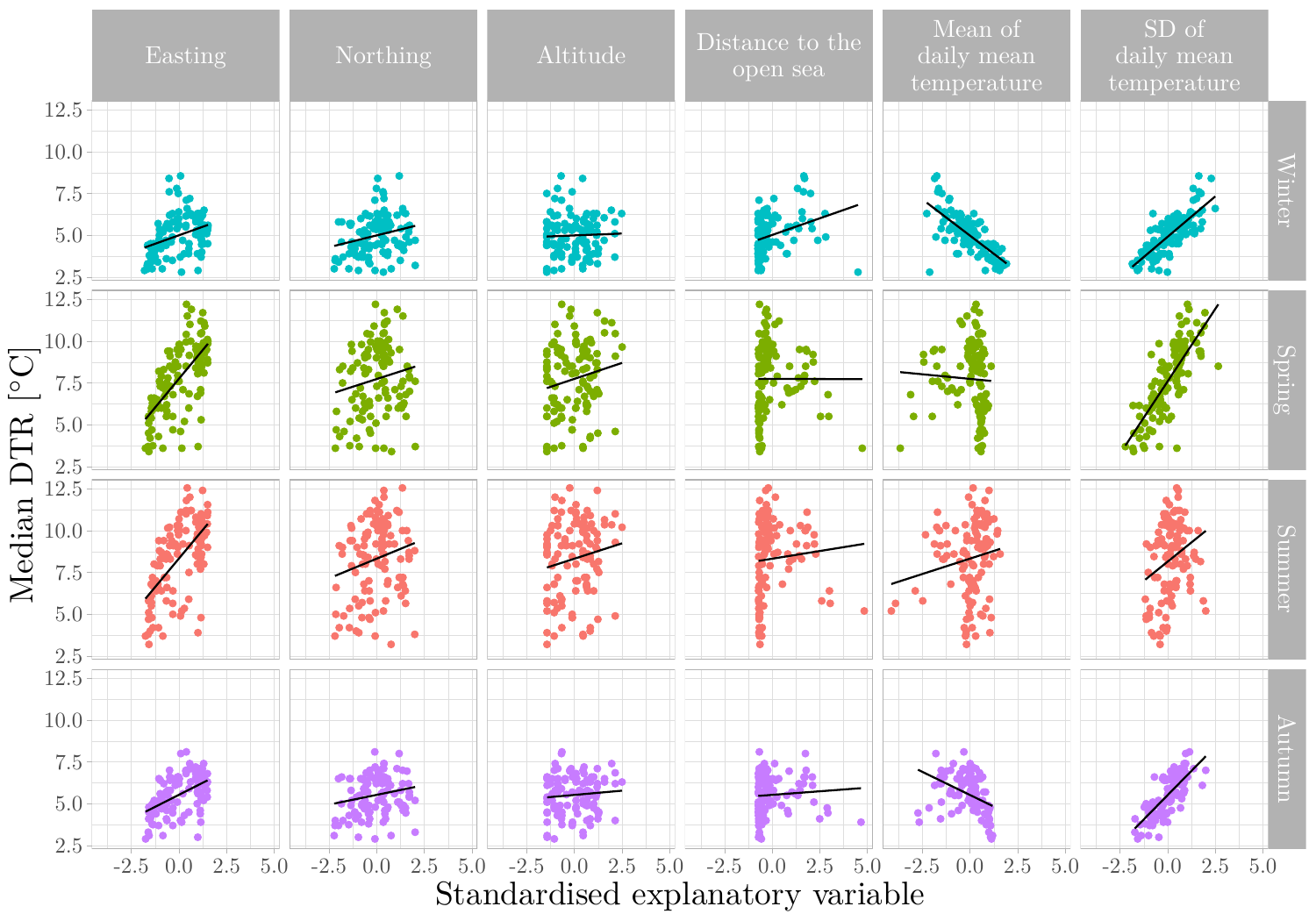}}
  \caption{Linear relationships between the six explanatory variables and
    the median of diurnal temperature range (DTR) at the 112 weather stations in our data set.
    All explanatory variables are standardised
  to have zero mean and a standard deviation of one.}
  \label{fig:covariates}
\end{figure}

To explore the relation between the explanatory variables and diurnal
temperature range, we fit a simple linear model with different quantiles of diurnal
temperature range against standardised versions of the explanatory variables. These linear model fits are
presented in Figure~\ref{fig:covariates} for the median of diurnal temperature range. Most of the
estimated trends are significant at the \(5\%\)-level. Especially for the mean and variance of
historical daily mean temperature, there is a strong linear relationship during
winter and autumn. Similar trends are found for all other examined quantiles.

\section{Models}
\label{sec:models}

\subsection{Marginal modelling with the FPL distribution}
\label{sec:fpld}
We assume that daily minimum and maximum
temperature can be described well by the generalised Pareto distribution. The Generalised Pareto
distribution is a common model
for extreme 
observations \citep[e.g.][]{coles2001introduction}, and it is often used for modelling extreme
temperature \citep[e.g.][]{castro-camilo21_bayes_space_time_gap_fillin,davison19_spatial_extrem}. Additionally,
\citet{rohrbeck21_spatio_tempor_model_red_sea} model daily temperature in the Red Sea by
assuming that both the upper and lower tails of daily temperature can be modelled using the generalised Pareto distribution, and
\citet{stein21_param_model_distr_when_inter} proposes to model climatological
phenomena using parametric distributions that behaves like the generalised Pareto distribution in both tails, and uses this
approach to model daily average temperature near Calgary during winter.
The generalised Pareto distribution can be described through its quantile function \citep[e.g.][]{hosking1987}
\begin{equation*}
  \label{eq:pareto_quantile}
  Q(p; \mu, \eta, \xi) = \inf \left\{x \in \mathbb{R}: p \leq F(x; \mu, \sigma, \xi)\right\} = \mu + \eta
  \begin{cases}
    \left(1 - (1 - p)^\xi\right) / \xi,& \xi \neq 0 \\
    - \log (1 - p),& \xi = 0,
  \end{cases}
\end{equation*}
where \(F\) is the cumulative distribution function, and $\mu$, $\eta$ and $\xi$ acts as the location, scale and
shape parameters, respectively.
Diurnal temperature range is equal to the difference between daily maximum and daily minimum
temperature, and can therefore be modelled as the difference between the quantiles of two generalised Pareto distributions.
We model maximum daily temperature as some quantile of a generalised Pareto distribution, while minimum
daily temperature is modelled as some quantile of a reflected generalised Pareto distribution.
If the random variable $X$ has a quantile function $Q_X(p)$, then $-X$ has the
quantile function $Q_{-X}(p) = -Q_X(1 - p)$. This results in an expression for the diurnal
temperature range:
\begin{equation*}
  Q_{\text{range}} = Q_{\text{max}}(p_1; \mu_1, \eta_1, \xi_1) -
    Q_{\text{min}}(p_2; \mu_2, \eta_2, \xi_2)
    = \mu_1 + \frac{\eta_1}{\xi_1} (1 - (1 - p_1)^{\xi_1}) + \mu_2 + \frac{\eta_2}{\xi_2} (1 - p_2^{\xi_2}).
\end{equation*}
In the case of \(\xi_1 = 0\) or \(\xi_2 = 0\) the expression is simplified, since
\(\lim_{\lambda \rightarrow 0} (p^\lambda - 1) / \lambda = \log p\).
We reparametrise by setting \(p_1 = p\), \(p_2 = p^a\), \(\xi_2^* = a
\xi_2\) and \(\eta_2^* = a \eta_2\), where \(a = \log p_2 / \log p\). This gives
\begin{equation*}
  Q_{\text{range}}(p) = \mu_1 + \frac{\eta_1}{\xi_1} (1 - (1 - p)^{\xi_1}) + \mu_2 + \frac{\eta_2^*}{\xi_2^*} (1 - p^{\xi_2^*}).
\end{equation*}
Further manipulation of the expression yields
\begin{equation*}
  Q_{\text{range}}(p) = (\mu_1 + \mu_2) + \frac{\eta_1 - \eta_2^*}{2} \left\{ \left(1 - \frac{\eta_1
        + \eta_2^*}{\eta_1 - \eta_2^*}\right) \frac{p^{\xi_2^*} - 1}{\xi_2^*} - \left(1 +
      \frac{\eta_1 + \eta_2^*}{\eta_1 - \eta_2^*}\right) \frac{(1 - p)^{\xi_1} - 1}{\xi_1}\right\},
\end{equation*}
which is equal to the quantile function of a five-parameter lambda (FPL)
distribution \citep{gilchrist2000},
\begin{equation}
  \label{eq:fpld}
  Q(p; \bm{\lambda}) = \lambda_1 +
  \frac{\lambda_2}{2} \left\{ (1 - \lambda_3) \frac{p^{\lambda_4} -
      1}{\lambda_4} - (1 + \lambda_3) \frac{{(1 - p)}^{\lambda_5} -
      1}{\lambda_5} \right\},\quad \lambda_2 > 0, \lambda_3 \in [-1, 1],
\end{equation}
Consequently, we expect the FPL distribution to be a suitable model for diurnal temperature range.

No analytic expression for the probability density function or cumulative
distribution function of the FPL distribution exists, although the density for a given
$p$ can be obtained as the reciprocal of the quantile derivative,
$(\mathrm{d}Q(p; \bm{\lambda})/\mathrm{d}p)^{-1}$.
From the quantile
function~\eqref{eq:fpld} of the FPL distribution it is clear that $\lambda_1$ acts as 
a location parameter and
$\lambda_2$ as a scale parameter of the distribution.
We notice that \(\lambda_4 = \xi_2^*\) and \(\lambda_5 = \xi_1\), which means that these two
parameters control the behaviour of the left and right tails, respectively.
The final parameter $\lambda_3$ acts as a weight between the two tails.

The support of the FPL distribution can be both finite and infinite. This makes the
distribution flexible for modelling a variety of different phenomena. The
support is given by
\begin{equation*} \label{eq:fpld_support} \left[Q(0; \bm{\lambda}), Q(1,
    \bm{\lambda})\right] = \lambda_1 + \frac{\lambda_2}{2}
  % \left\{
  \begin{cases}
    [- \frac{1 - \lambda_3}{\lambda_4}, \frac{1 + \lambda_3}{\lambda_5}],
    & \lambda_4, \lambda_5 > 0 \\
    [- \frac{1 - \lambda_3}{\lambda_4},
    \infty), & \lambda_4 > 0, \lambda_5 \leqslant 0 \\
    (-\infty, \frac{1 + \lambda_3}{\lambda_5}], &\lambda_4 \leqslant 0,
    \lambda_5 > 0
  \end{cases}.
  % \right.
\end{equation*}
Diurnal temperature range is always positive. In order to ensure a positive
support for the FPL distribution, one must enforce the inequality-constraints
\begin{equation}
  \label{eq:FPLD_restrictions}
  \lambda_1 - \frac{\lambda_2 (1 - \lambda_3)}{2 \lambda_4} > 0,
  \qquad  \lambda_4 > 0.
\end{equation}

The parametrisation in~\eqref{eq:fpld} is intuitive in the sense
that it stems from the combination of two generalised Pareto distributions. However, for performing numerical
parameter estimation, other representations are more appropriate.
The location \(\lambda_1\) and scale \(\lambda_2\) are not clearly linked to any central moments or 
quantiles of the FPL distribution. We thus propose a reparametrisation scheme with a new location parameter that is equal to the
median of the FPL distribution, and a new scale parameter that is equal to the inter-quartile range of the FPL distribution,
\begin{equation}
  \label{eq:reparam1}
  \begin{aligned}
    \lambda_1^* &= Q(0.5; \bm{\lambda}) = \lambda_1 + \frac{\lambda_2}{2} \left\{ (1 - \lambda_3) \frac{{0.5}^{\lambda_4} -
        1}{\lambda_4} - (1 + \lambda_3) \frac{{0.5}^{\lambda_5} -
        1}{\lambda_5} \right\}, \\
    \lambda_2^* &= Q(0.75; \bm{\lambda}) - Q(0.25; \bm{\lambda}) = \frac{\lambda_2}{2}
    \left\{\frac{1 - \lambda_3}{\lambda_4}\left(0.75^{\lambda_4} - 0.25^{\lambda_4}\right) + \frac{1 +
        \lambda_3}{\lambda_5}\left(0.75^{\lambda_5} - 0.25^{\lambda_5}\right) \right\}.
  \end{aligned}
\end{equation}
The new parameter vector is denoted \(\bm{\lambda}^* = (\lambda_1^*, \lambda_2^*,
\lambda_3, \lambda_4, \lambda_5)\).
In order to simplify parameter constraints during any numerical estimation procedures, we further introduce
the reparametrisation
\begin{equation}
  \label{eq:reparam2}
  \tilde{\lambda}_1 = \lambda_1^*,\quad
  \tilde{\lambda}_2 = \log\left(e^{\lambda_2^*} - 1\right),\quad
  \tilde{\lambda}_3 = \log\left(\frac{1 - \lambda_3}{1 + \lambda_3}\right),\quad
  \tilde{\lambda}_4 = \log\left(e^{\lambda_4} - 1\right),\quad
  \tilde{\lambda}_5 = \log\left(e^{\lambda_5 + 0.5} - 1\right),\quad
\end{equation}
This results in an unconstrained parameter vector \(\tilde{\bm{\lambda}} \in \mathbb{R}^5\) and guarantees
that \(\lambda_2 > 0\), \(\lambda_3 \in (-1, 1)\) and \(\lambda_4 > 0\). We also restrict
\(\lambda_5\) to the interval \((-0.5, \infty)\), as this guarantees a finite mean and variance for
the FPL distribution \citep[e.g.][]{coles2001introduction,tarsitano2010}. Exploratory data analysis (results not shown) finds that the right tail parameter in the
diurnal temperature range distribution tends to be considerably larger than
-0.5. \citet{davison19_spatial_extrem} model extreme temperatures in Spain with the generalised Pareto distribution and find that the
tail parameter, which we denote by \(\lambda_5\), is approximately equal to 0.4.
\citet{castro-camilo21_bayes_space_time_gap_fillin,rohrbeck21_spatio_tempor_model_red_sea}
model Red Sea temperatures with the generalised Pareto distribution and find that the tail parameter is larger than
\(-0.1\). \citet{osullivan20_bayes_spatial_extrem_value_analy} model temperature extremes in Dublin
and find that the posterior median of the tail parameter is larger than \(0.1\). Based on these results
and our exploratory data analysis, we are confident
that the restriction of \(\lambda_5 > -0.5\) should not lead to any loss in model performance.

The standard way of reparametrising a parameter \(\theta\) that is bounded away from zero is to set
\(\tilde{\theta} = \log \theta\). However, if \(\theta\) attains a large value, a small error in
the estimate for \(\tilde{\theta}\) leads to a considerable error in the estimate for \(\theta\).
The function \(g(x) = \log(e^x - 1)\) has the property that \(g(x) \approx \log x\) for small \(x\)
and \(g(x) \approx x\) for large \(x\). This allows us to constrain the FPL parameters without
risking large reparametrisation instability because of an exponential relation between \(\bm{\lambda}\)
and \(\tilde{\bm{\lambda}}\).

The reparametrisation to \(\tilde{\bm{\lambda}}\) eases numerical inference methods, and is used whenever
we perform parameter estimation. However, the \(\bm{\lambda}^*\)
parametrisation is more intuitive, and is therefore primarily used when describing our methods.

\subsection{Distributional quantile regression}
\label{sec:quantile_regression}

We wish to model the marginal distribution of diurnal temperature range at locations without available
temperature observations, using a regression model with explanatory variables. As described in
Section~\ref{sec:data}, the distribution of diurnal temperature range is very rich. Many of its
distributional properties seem to vary in space, and between seasons. Thus, it does not seem
good enough to use e.g.\ a generalised linear model (GLM) where we only allow the mean and
variance of diurnal temperature range to vary in space. We suggest that one should apply a
distributional regression model, where the entire distribution function is allowed to vary in
space, \citep[e.g.][]{henzi20_isoton_distr_regres, schlosser19_distr_regres_fores_probab_precip,klein15_bayes_struc_addit_distr_regres}
for modelling diurnal temperature range. One way of performing distributional regression
is to apply a latent Gaussian model with an FPL likelihood, such that
all five FPL parameters are modelled as a linear combination of explanatory variables and
Gaussian white noise. However, this leads to an unnecessarily complex model. Additionally, as described in
Section~\ref{sec:simulation-study}, the flexibility of the five FPL parameters might lead to
something similar to identifiability problems, that can be problematic when we perform regression directly
on the parameters and not on the distribution itself. Here, we propose a novel distributional regression
model for diurnal temperature range which is based on combining quantile regression and marginal
modelling with the FPL distribution, and we use this for modelling the marginal distribution of diurnal
temperature range at locations with no temperature observations.

We first assume that any quantile in the distribution of diurnal temperature range
can be modelled as a linear combination of explanatory variables.
Let \(y_i(\bm{s})\) be observation \(i\) of diurnal temperature range at location \(\bm{s} \in
\mathcal{S}\), where \(\mathcal{S}\) is the given study area. For any probability \(p \in (0, 1)\),
we assume that the diurnal temperature range can be modelled as
\begin{equation}
  \label{eq:regression model} 
  y_i(\bm{s}) = {\textbf{x}(\bm{s})}^T \bm{\beta}_{p} + \epsilon_{i, p}(\bm{s}),\quad i = 1,
  \ldots, n(\bm{s}),
\end{equation}
with explanatory variables \(\textbf{x}(\bm{s})\) and regression coefficients \(\bm{\beta}_p\). The
error terms \(\epsilon_{i, p}(\bm{s})\) are assumed to be independent and distributed such that
\(P(\epsilon_{i, p}(\bm{s}) \leq 0) = p\) for all \(\bm{s} \in
\mathcal{S}\) and \(i = 1, \ldots, n(\bm{s})\) \citep{koenker2005quantileRegression}.
We are now able to estimate quantiles \(q_p(\bm{s}) = \textbf{x}(\bm{s})^T \bm{\beta}_p\) of diurnal
temperature range for any \(p \in (0, 1)\), at any location \(\bm{s}\) with available explanatory
variables \(\textbf{x}(\bm{s})\).

In order to turn this into a distributional regression model, we
further propose to treat all \(q_p(\bm{s})\) as quantiles of the FPL distribution at location
\(\bm s\).
The only necessary assumption for performing quantile regression is that \(P(y_i(\bm{s}) \leq
\textbf{x}(\bm{s})^T \bm{\beta}_p) = p\), and there need not
be any disagreements between this and the assumption that the marginal distribution at any
location \(\bm s\) is the FPL distribution. Consequently, we model the distribution of diurnal
temperature range at location \(\bm s\) using the FPL distribution with the parameters \(\bm \lambda^*\) that
minimise the distance between \(q_p(\bm s)\) and the quantile function \(Q(p; \bm \lambda^*)\) of
the FPL distribution \eqref{eq:fpld}.
With this approach, we are able to describe the distribution of diurnal temperature range
everywhere, using a parametric model. This makes it easier to interpret the distributional
properties of diurnal temperature range than when we only use the semi-parametric quantile
regression model.

A common problem with quantile regression is that the different estimated quantile models may cross,
such that the estimator for  $q_{p_i}(\bm{s})$ is larger than the estimator for $q_{p_j}(\bm{s})$
for \(p_j > p_i\) \citep{rodrigues2017regression,cannon18_non_cross_nonlin_regres_quant,bondell10_noncr_quant_regres_curve}.
However, by first performing quantile regression and then fitting the FPL distribution 
to the regression quantiles, this problem is easily fixed, as the FPL quantile function always is
monotonic increasing. Consequently, there is no need to implement complicated
quantile regression methods that ensure non-crossing quantiles, and we can base our modelling on the
fast and simple regression model where each quantile is modelled separately.

For simplicity, the distributional quantile regression model is referred to as the regression model for the
remainder of the paper.

\section{Inference}
\label{sec:inference}

\subsection{Parameter estimation for the FPL distribution}
\label{sec:univariate-methods}

We present three marginal parameter estimation methods for the FPL distribution: the method of quantiles,
maximum likelihood estimation and the starship method.

\subsubsection{The method of quantiles}

The method of quantiles is an estimation method similar to
the better known method of moments, in which the distance between
quantiles of a parametric distribution and empirical quantiles from
observed data is minimised.
Let $y_1, \ldots, y_n$ be independent and identically FPL distributed random
variables with quantile function $Q(p; \bm{\lambda}^*)$ and parameters
  \(\bm{\lambda}^*\). A set of \(m\) empirical quantiles
$\widehat{Q}(p_i) = \widehat{q}_{p_i},\ i = 1, 2, \ldots, m,$ are constructed from the observations $\bm{y}$.
The method of quantiles estimator
for $\bm{\lambda}^*$, is found by minimising the
absolute distance% between $Q(p; \bm{\lambda}^*)$ and $\widehat{q}_p$,
\begin{equation}
  \label{eq:mq}
  L(\bm{\lambda}^*) = \sum_{i = 1}^m \left|\widehat{q}_{p_i} - Q(p_i; \bm{\lambda}^*)\right|.
\end{equation}

There do not exist any straightforward expressions for the probability density or the cumulative
probability function of the FPL distribution. However, a simple and closed-form expression exists for its quantile
function. This can make it more natural to perform parameter estimation based on quantile matching
instead of e.g.\ likelihood-based estimation methods.
In addition, \citet{bignozzi2018method_of_quantiles} state that
parameter estimation methods based on quantile matching can be
preferable when distributions are heavy-tailed or their support varies
with the parameters. Both of these conditions hold for the FPL distribution.
Additionally, \citet{bhatti2018pareto}
find that the method of quantiles outperforms both the method of
moments and maximum likelihood estimation for parameter
estimation under the Pareto distribution. As the FPL distribution can be described
as the difference between two Pareto distributions, it should share some of
the same properties.
\citet{tarsitano2005} applies the method of quantiles for parameter
estimation with the FPL distribution, using only five quantiles. He concludes
that the method has several advantages, while a theoretical
justification for the choice of quantiles is lacking.
For a large set of observations
$y_{(1)} \leqslant y_{(2)} \leqslant \ldots y_{(n)}$, the distribution
function of $y_{(i)}$ is close to the empirical distribution function
$\widehat{F}(y_{(i)}) = (i - 0.5) / n$. Consequently, for a large set of
$n$ observations, we perform the method of quantiles by setting
$p_i = (i - 0.5)/n$ and $\widehat{q}_{p_i} = y_{(i)}$ for
$i = 1, 2, \ldots, n$, thus avoiding the issue of which quantiles to select. This is somewhat similar
to the method of least absolute deviations by \citet{tarsitano2010}.
\citet{koenker2005quantileRegression} has shown that
the method of quantiles estimator is consistent,
provided that all estimated quantiles $\widehat{q}$ are consistent.
Note that the order statistics of \(\bm{y}\) are dependent. This must be taken into account if
one attempts to compute the variance of the estimator for \(\bm{\lambda}^*\).

A weakness of the method of quantiles is that it  might return a parameter estimator
$\widehat{\bm{\lambda}^*}$ such that
$Q(0; \bm{\widehat{\lambda}}^*) > y_{(1)}$ or
$Q(1; \widehat{\bm{\lambda}}^*) < y_{(n)}$. This problem is addressed by introducing
inequality constraints when minimising the loss function in \eqref{eq:mq}, demanding
that $Q(0; \bm{\lambda}^*) < y_{(1)}$ and $y_{(n)} < Q(1; \bm{\lambda}^*)$. In order to guarantee a
positive support, the inequality constraints from \eqref{eq:FPLD_restrictions} can also be
enforced. Note that \(\lambda_4 > 0\) is automatically enforced by optimising over the
\(\tilde{\bm{\lambda}}\) parametrisation.
The positive support constraint is slightly relaxed by only demanding that \(Q(10^{-4};
\bm{\lambda}^*) > 0\), as we find that this can considerably improve the model fit in certain cases.
These constraints are enforced by performing numerical
optimisation using an augmented Lagrangian formulation \citep[e.g.][]{nocedal06_numer},
implemented within the \texttt{R} package \texttt{nloptr}
\citep{johnson20_nlopt,birgin08_improv_ultim_conver_augmen_lagran_method}.
Closed-form expressions are available both for the quantile loss function and for its gradient.
However, in practice we find that the method of quantiles performs better when not including
gradient information in the optimiser. Consequently, we
minimise the augmented Lagrangian using the derivative-free Nelder-Mead algorithm
\citep{nelder65_simpl_method_funct_minim}, also implemented in the \texttt{nloptr} package.

Due to the flexibility of the FPL distribution, the quantile loss \eqref{eq:mq} proves to be difficult to
minimise without a good initial value for the FPL parameters. We
utilise the connection between \((\lambda_1^*, \lambda_2^*)\) and the quantiles of
the FPL distribution by setting the initial values equal to the empirical median and inter-quartile range of
\(\bm{y}\), respectively. Initial values for
the remaining three parameters are selected using a quick grid search. We compute the
quantile loss function for all combinations of \(\lambda_3 \in \{-0.5, -0.25, 0, 0.25, 0.5\}\),
\(\lambda_4 \in \{0.1, 0.2, 0.4, 0.8, 1, 1.5\}\) and \(\lambda_5 \in \{-0.4, -0.1, 0.1, 0.2, 0.4,
0.8, 1, 1.5\}\) and select the combination of parameters that minimises it as initial values.

\subsubsection{Maximum likelihood estimation}

No closed-form expression exists for the cumulative distribution function of the FPL distribution. However, as
mentioned in Section~\ref{sec:fpld}, 
for a given probability \(p\), the probability density function of the FPL distribution can be obtained as the
reciprocal of the quantile derivative
\begin{equation*}
  f(y; \bm{\lambda}) = \frac{2}{\lambda_2}\left\{(1 - \lambda_3)p^{\lambda_4 - 1} + (1 + \lambda_3)(1
    - p)^{\lambda_5 - 1}\right\}^{-1},
\end{equation*}
with \(p = F(y; \bm{\lambda})\). A numerical approximation for the cumulative distribution function
of the FPL distribution is available from the \texttt{R} package \texttt{gld} \citep{king20}. For \(\bm{y} =
(y_1, \ldots, y_n)^T\) this gives rise to the log-likelihood 
\begin{equation*}
  \ell(\bm{\lambda}; \bm{y}) = n \log 2 - n \log \lambda_2 - \sum_{i = 1}^n \log \left\{(1 -
    \lambda_3)p(y_i)^{\lambda_4 - 1} + (1 + \lambda_3)(1 - p(y_i))^{\lambda_5 - 1}\right\}.
\end{equation*}
A straightforward expression for the gradient of the log-likelihood cannot be provided, as it requires
computing the derivative of \(F(y; \bm{\lambda})\) with respect to \(\bm{\lambda}\).
Maximisation of the log-likelihood is performed using the same grid-search and augmented
Lagrangian as for the method of quantiles, where we include the same inequality constraint to guarantee
a positive support.

\subsubsection{The starship method}
A straightforward way of modelling with the FPL distribution is to use the already implemented functions in the
\texttt{R} package \texttt{gld} \citep{king20}. This package is mostly focused on the
four-parameter
generalised lambda distribution, but it also includes one inference method for the FPL distribution, namely the
starship method \citep{king99_theor_method,owen88_stars}. The starship method is based on
the fact that if \(\bm{y}\) has distribution function \(F(\cdot; \bm{\lambda}^*)\), the
transformed variable \(\bm{u} = F(\bm{y}; \bm{\lambda}^*)\) has a uniform distribution. Parameters can
therefore be estimated by minimising any goodness-of-fit statistic between the uniform distribution
and \(F(\bm{y}; \bm{\lambda}^*)\).

The starship suffers from the same problems as the maximum likelihood estimator,
namely that no closed-form expression exists for the cumulative distribution function of the FPL distribution,
which therefore has to be numerically approximated. The implemented method in the \texttt{gld} package
performs minimisation using the Anderson-Darling statistic. However, the \texttt{gld} implementation of
the starship method is very computationally inefficient, so in order to compare it with the previously
described methods for large amounts of data, we implement our own version of the starship, based on the \texttt{gld}
implementation. This implementation performs minimisation of the Anderson-Darling statistic using
the same optimisation approach as in the precious methods. On a sample of \(10^4\)
observations, our version of the starship estimates \(\bm{\lambda}^*\) in approximately 50
seconds, while the \texttt{gld} implementation uses approximately 330 seconds. The two implementations
seem to perform equally well numerically.

\subsection{Parameter estimation in the regression model}
\label{sec:qr_estimation}

\begin{figure}
  \centerline{\includegraphics[width=.8\linewidth]{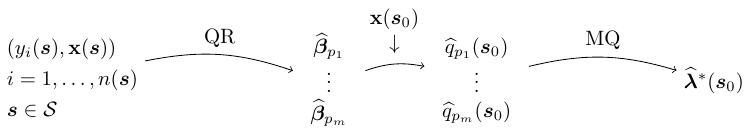}}
  \caption{Diagram of the regression model. First, quantile regression (QR)
    is performed for modelling different quantiles of diurnal temperature range $y_i(\bm{s})$ given
    the explanatory variables $\textbf{x}(\bm{s})$. Using the quantile regression
    model, the conditional quantiles
    $\widehat{q}_{p_1}(\bm{s}_0), \widehat q_{p_2}(\bm s_0) \ldots, \widehat{q}_{p_m}(\bm{s}_0)$ are estimated, given the
    explanatory variables $\textbf{x}(\bm{s}_0)$ at a specific location. Then, using the
    method of quantiles (MQ), the FPL distribution is fitted to the conditional
    quantiles, resulting in the estimator $\widehat{\bm{\lambda}}^*(\bm{s}_0)$ for the FPL
    parameters.}
  \label{fig:scheme}
\end{figure}

Inference for the regression model is divided into two steps. The estimation
procedure is illustrated in Figure~\ref{fig:scheme}. First, quantile
regression is performed using the \texttt{R} package \texttt{quantreg} \citep{quantreg},
separately for each of the probabilities $p_i = i / 100,\ i = 1, 2, \ldots, 99$. Then, at any location
\(\bm{s}_0\) with available explanatory variables where we wish to model diurnal temperature range, 
we estimate the conditional quantiles $\widehat{q}_{p_i}(\bm{s}_0)$.
The quantile function of the FPL distribution is then fitted to the 99 estimated quantiles
\(\widehat{q}_{p_1}(\bm{s}_0), \widehat q_{p_2}(\bm s_0) \ldots, \widehat{q}_{p_{99}}(\bm{s}_0)\) using a marginal parameter
estimation method. There are no good ways of extending the maximum likelihood or starship method to
fit a quantile function to a set of quantiles. However, the method of quantiles is perfect for this
kind of parameter estimation problem. Consequently,
the method of quantiles is used for fitting the FPL distribution to the 99 estimated quantiles,
resulting in an estimator $\widehat{\bm{\lambda}}^*(\bm{s}_0)$ for the FPL parameters at \(\bm s_0\).
All 99 quantiles are modelled independently of each other, meaning that numerical inference can be executed in
parallel. Fitting the FPL distribution to estimated quantiles at different locations is also an
independent operation that can be performed in parallel. The proposed regression model is therefore
highly parallelisable.

\subsection{Model evaluation}
\label{sec:evaluation}

Model performance is evaluated using the continuous ranked probability score
\citep[CRPS][]{gneiting07_stric_proper_scorin_rules_predic_estim,matheson76_scorin_rules_contin_probab_distr}. Given
a forecast distribution \(F\) and an observation \(y\), the CRPS is equal to
\begin{equation}
  \label{eq:crps}
  S(F, y) = \int_{-\infty}^\infty (F(t) - I(t \geq y))^2 \text{ d}t = 2 \int_0^1 \rho_p(y -
  F^{-1}(p)) \text{ d}p,
\end{equation}
with \(\rho_p(u) = pu - I(u < 0) u\), where \(I(\cdot)\) is an indicator function.
This is a strictly proper scoring rule, meaning that if \(y\) has distribution function \(G\), then \(E[S(G, y)] \leq E[S(F,
y)]\) for all forecast distributions \(F\), with equality only if \(F = G\). Due to scarce usage of
the FPL distribution in the literature, to the best of our knowledge, a closed-form expression for the CRPS with
an FPL forecast distribution has not yet been provided. Consequently, we derive the expression
for the CRPS with an FPL forecast distribution. When modelling diurnal temperature range with the FPL distribution, the
quantile formulation of the CRPS is especially useful.
Given an FPL forecast distribution \(F\) with quantile function \(Q\), the CRPS can be expressed as
\begin{equation*}
  S(F, y) = 2\int_0^1 \rho_p(y - Q(p)) \text{ d}p = y (2 F(y) - 1) - 2 \int_0^1 p Q(p) \text{ d}p + 2
  \int_{F(y)}^1 Q(p) \text{ d}p.
\end{equation*}
Both of these integrals are fairly straight-forward to compute with the FPL quantile function, as
they are simply polynomials in \(p\). Solving the first integral yields
\begin{equation*}
  \begin{aligned}
    &\int_{F(y)}^1 Q(p) \text{ d} p = \\
    &\qquad (1 - F(y))\lambda_1 + \frac{\lambda_2}{2} \left(
    (1 - \lambda_3) \left(\frac{1 - F(y)^{\lambda_4 + 1}}{\lambda_4 (\lambda_4 + 1)} - \frac{1 -
        F(y)}{\lambda_4}\right) -
    (1 + \lambda_3) \left(\frac{(1 - F(y))^{\lambda_5 + 1}}{\lambda_5 (\lambda_5 + 1)} - \frac{1 - F(y)}{\lambda_5}\right)\right),
  \end{aligned}
\end{equation*}
while solving the second integral yields
\begin{equation*}
    \int_0^1 p Q(p) \text{ d} p =
    \frac{\lambda_1}{2} + \frac{\lambda_2}{2}\left(-(1 - \lambda_3)\frac{1}{2 (\lambda_4 + 2)} + (1
      + \lambda_3)\frac{\lambda_5 + 3}{2 (\lambda_5 + 1) (\lambda_5 + 2)}\right).
\end{equation*}
Thus, by numerically approximating \(F(y)\) we are able to estimate the CRPS of the FPL distribution.
In the special case of \(\lambda_4 = 0\) or \(\lambda_5 = 0\), the integrals are solved by using that
\begin{equation*}
  \begin{aligned}
  &\ \ \int \log p \text{ d} p = p (\log p - 1), \qquad \qquad \ \  &&\ \ \int \log (1 - p) \text{ d} p = (p - 1) \log (1 - p)
  - p, \\
  &\int p \log p \text{ d} p = \frac{1}{2} p^2 (2 \log p - 1), \qquad &&\int p \log(1 - p) \text{ d}p =
  \frac{1}{4} \left(2 (p^2 - 1) \log(1 - p) - p (p + 2)\right).
  \end{aligned}
\end{equation*}

The CRPS can be used for comparing competing forecasts. Given two forecasts \(F_1\) and \(F_2\),
and observations \(\bm{y} = (y_1, \ldots y_n)^T\), we can compute the mean CRPS
\[S(F, \bm{y}) = \frac{1}{n} \sum_{i = 1}^n S(F, y_i),\]
and choose the forecast with the lowest mean CRPS.
However, the mean CRPS in itself does not provide much information about the goodness of fit of a
forecast. Given a forecast \(F\) and observations \(\bm{y}\) with unknown distribution function
\(G\), there is no way of knowing if there is a large difference between \(S(F, \bm{y})\) and
\(S(G, \bm{y})\), since \(G\) is unknown.
Thus, we also evaluate model performance by studying quantile-quantile-plots (QQ-plots)
and the probability integral transform (PIT).

If a random variable $y$ has distribution function $F$, then the transformed
variable $u = F(y)$ is uniformly distributed between zero and one. Given a forecast
distribution \(F\) and observations \(\bm{y}\) one can therefore examine
deviations between the distribution of \(\bm{u} = F(\bm{y})\) and the standard uniform
distribution. \citet{heinrich20_multiv_postp_method_high_dimen} propose to evaluate model fit by examining the first two
moments of \(\bm{u}\). Denote the error in the first moment as $e_\mu = E(\bm{u}) - 0.5$
and the error in the second central moment as $e_\sigma = \text{SD}(\bm{u}) - 1 /
\sqrt{12}$. It follows that $e_\mu < 0$ indicates a positive bias and $e_\mu > 0$ indicates a
negative bias. If $e_\sigma < 0$, the forecast distribution \(F\) is overdispersive, and
  if $e_\sigma > 0$, it is underdispersive.

\section{Simulation study}
\label{sec:simulation-study}

\subsection{Setup}

Simulation studies are performed to compare the different parameter estimation methods for the
FPL distribution. We draw 500 random sets of FPL parameters \(\bm{\lambda}^*\). These are sampled such that they
are of approximately the same magnitude as the estimated FPL parameters for diurnal temperature
range in Section~\ref{sec:results} (see Table~\ref{tab:param-table}), while also ensuring that we
have a positive and wide enough support. The exact sampling scheme is given in
Algorithm~\ref{alg:simulation}, with \(\mathcal{N}(\mu, \sigma^2)\) denoting a Gaussian distribution
with mean \(\mu\) and variance \(\sigma^2\) and \(U(a, b)\) denoting a uniform distribution with
limits \(a\) and \(b\).
For each set of FPL parameters we then sample \(n\) realisations from the
FPL(\(\bm{\lambda}^*\)) distribution with \(n = 2^i\) for \(i = 7, 8, \ldots, 14\).
The parameter estimation methods described in Section~\ref{sec:univariate-methods} are then applied for estimating
\(\bm{\lambda}^*\). We evaluate the
overall fit to data and the ability of recovering the true parameter values. Overall model fit to
data is evaluated using the CRPS. Since the true value of \(\bm{\lambda}^*\) is known, we can compute
the skill score \( 1 - \text{CRPS}(F, G) / \text{CRPS}(F, y)\), where \(F\) is our estimated distribution, \(G\) is the
correct distribution and CRPS\((F, G)\) is the expected value of the CRPS with respect to \(G\). Thus,
a perfect forecast gives a skill score of zero, while all other forecasts give a skill score larger
than zero and smaller than one. The ability to recovery the true values of \(\bm{\lambda}^*\) is evaluated using the mean
square error (MSE) between the true parameters \(\bm{\lambda}^*\) and the estimated parameters
\(\hat{\bm{\lambda}}^* \), over all 500 repetitions,
\begin{equation*}
  \text{MSE}(\bm \lambda^*, \bm \lambda) = \frac{1}{500} \sum_{i = 1}^{500} \frac{1}{5} \sum_{j =
    1}^5 \left({\lambda_j^*}^{(i)} - \lambda_j^{(i)}\right)^2,
\end{equation*}
where \(\bm \lambda^{(i)} = (\lambda_1^{(i)}, \ldots \lambda_5^{(i)})^T\) are the true FPL
parameters in simulation number \(i\) out of 500, and \({\bm{\lambda^*}}^{(i)}\) are the
corresponding estimated parameters.

\begin{algorithm}
  \caption{Sampling \(\bm{\lambda}^*\)}
  \label{alg:simulation}
  \begin{algorithmic}
    \While{TRUE}
    \State Sample \(\lambda_1^* \sim \mathcal{N}(5, 3^2)\)
    \State Sample \(\lambda_2^* \sim U(1.5, 8)\)
    \State Sample \(\lambda_3 \sim U(-0.9, 0.9)\)
    \State Sample \(\lambda_4 \sim U(0.01, 0.9)\)
    \State Sample \(\lambda_5 \sim U(-0.3, 0.7)\)
    \If {\(Q(0, \bm{\lambda}^*) > 0\)} \Comment{Ensure a positive support}
    \If {\(Q(1, \bm{\lambda}^*) - Q(0, \bm{\lambda}^*) > 1\)} \Comment{Ensure that the support does
      not become too compact}
    \State \textbf{break}
    \EndIf
    \EndIf
    \EndWhile
    \State \Return \(\bm{\lambda}^*\)
  \end{algorithmic}
\end{algorithm}

\subsection{Results}

Table~\ref{tab:univariate-simulation} displays the skill score, MSE and computation
time for all methods. As \(n\) grows, the computation times for the maximum likelihood and starship
methods grow considerably faster than the time for the method of quantiles. This happens because
the method of quantiles is based solely on analytical expressions, while the other two methods
require numerical estimation of the likelihood or distribution function of the FPL distribution. The method of
quantiles has a worse skill score for small to medium sample sizes and a slightly better skill score for
large sample sizes, whereas the starship method attains the worst skill score for large sample sizes.
Interestingly, the method of quantiles fails to recover
the correct FPL parameters, and has a much larger MSE than the other two methods. This demonstrates the
flexibility of the FPL distribution, as we are able to achieve a better CRPS using the ``wrong'' parameter
estimates. A closer examination of the estimated parameters finds that the large increase in MSE is caused
almost solely by too large estimates of \(\lambda_4\) and \(\lambda_5\). In some situations it seems
that a large increase in \(\lambda_4\) combined with a decrease in
\(\lambda_3\) yields almost no change in the overall shape of the FPL distribution. This makes sense, as increasing
\(\lambda_4\) leads to a thinner left tail, while decreasing \(\lambda_3\) places more weight on the
left tail. Similarly, a large increase in \(\lambda_5\) can be mitigated by increasing \(\lambda_3\).
When modelling diurnal temperature range, model fit is much more important than parameter
recovery, as the FPL model is merely an assumption, and possibly not the true underlying distribution
of diurnal temperature range. Thus, the low skill score and fast computation times of the
method of quantiles for large sample sizes make up for the fact that we seem to lose the ability to
always recover the true parameters.

\begin{table}
  \centering
  \renewcommand*{\arraystretch}{1.5}
  \caption{Mean skill scores, MSE and computation times for the method of quantiles (MQ), maximum
    likelihood (ML) and starship method, when performing parameter estimation 500 times on \(n\) samples drawn
    from an FPL distribution. The skill scores are multiplied by \(10^3\) to get more readable
    results. Computation times are reported on a 2.4 gHz computation server.} 
  \label{tab:univariate-simulation}
  \resizebox{\textwidth}{!}{%
    \begin{tabular}{llllllllll}
      \toprule
      & Method &  \(n = 2^{7}\) & \(n = 2^{8}\) & \(n = 2^{9}\) & \(n = 2^{10}\) & \(n = 2^{11}\) & \(n = 2^{12}\) & \(n = 2^{13}\) & \(n = 2^{14}\) \\
      \midrule
      Skill score (\(\cdot 10^{3}\)) & ML & \(6.12\) & \(3.18\) & \(1.48\) & \(0.73\) & \(0.40\) & \(0.21\) & \(0.11\) & \(0.09\) \\
      & MQ & \(6.42\) & \(3.33\) & \(1.58\) & \(0.81\) & \(0.43\) & \(0.21\) & \(0.10\) & \(0.06\) \\
      & starship & \(6.12\) & \(3.18\) & \(1.49\) & \(0.75\) & \(0.40\) & \(0.26\) & \(0.20\) & \(0.21\) \\
      \midrule
      MSE & ML & \( 0.17\) & \(  0.14\) & \(  0.06\) & \(0.03\) & \( 0.02\) & \(0.02\) & \(0.02\) & \(0.01\) \\
      & MQ & \(14.01\) & \(207.16\) & \(201.66\) & \(6.87\) & \(32.20\) & \(3.85\) & \(5.04\) & \(0.47\) \\
      & starship & \( 1.27\) & \(  0.35\) & \(  0.28\) & \(0.26\) & \( 0.15\) & \(0.02\) & \(0.02\) & \(0.03\) \\
      \midrule
      Time [s] & ML & \(0.6\) & \(0.9\) & \(1.5\) & \(3.0\) & \(5.6\) & \(11.0\) & \(22.6\) & \(46.4\) \\
      & MQ & \(0.2\) & \(0.2\) & \(0.3\) & \(0.3\) & \(0.4\) & \( 0.6\) & \( 1.1\) & \( 1.9\) \\
      & starship & \(0.8\) & \(1.1\) & \(1.8\) & \(3.3\) & \(6.5\) & \(14.1\) & \(30.4\) & \(63.9\)  \\
      \bottomrule
    \end{tabular}
  }
\end{table}

\section{Modelling diurnal temperature range in southern Norway}
\label{sec:results}

Diurnal temperature range in southern Norway is modelled separately for all seasons,
using our two proposed models.
Model calibration is evaluated using QQ-plots and the PIT. The marginal model is
compared against several competing models using the CRPS. The 
regression model is tested in a leave-one-out cross-validation study, and the results are
compared with the marginal model fits.

\subsection{Marginal modelling}
\label{sec:mq_results}

\begin{figure}
  \centerline{\includegraphics[width=.8\linewidth]{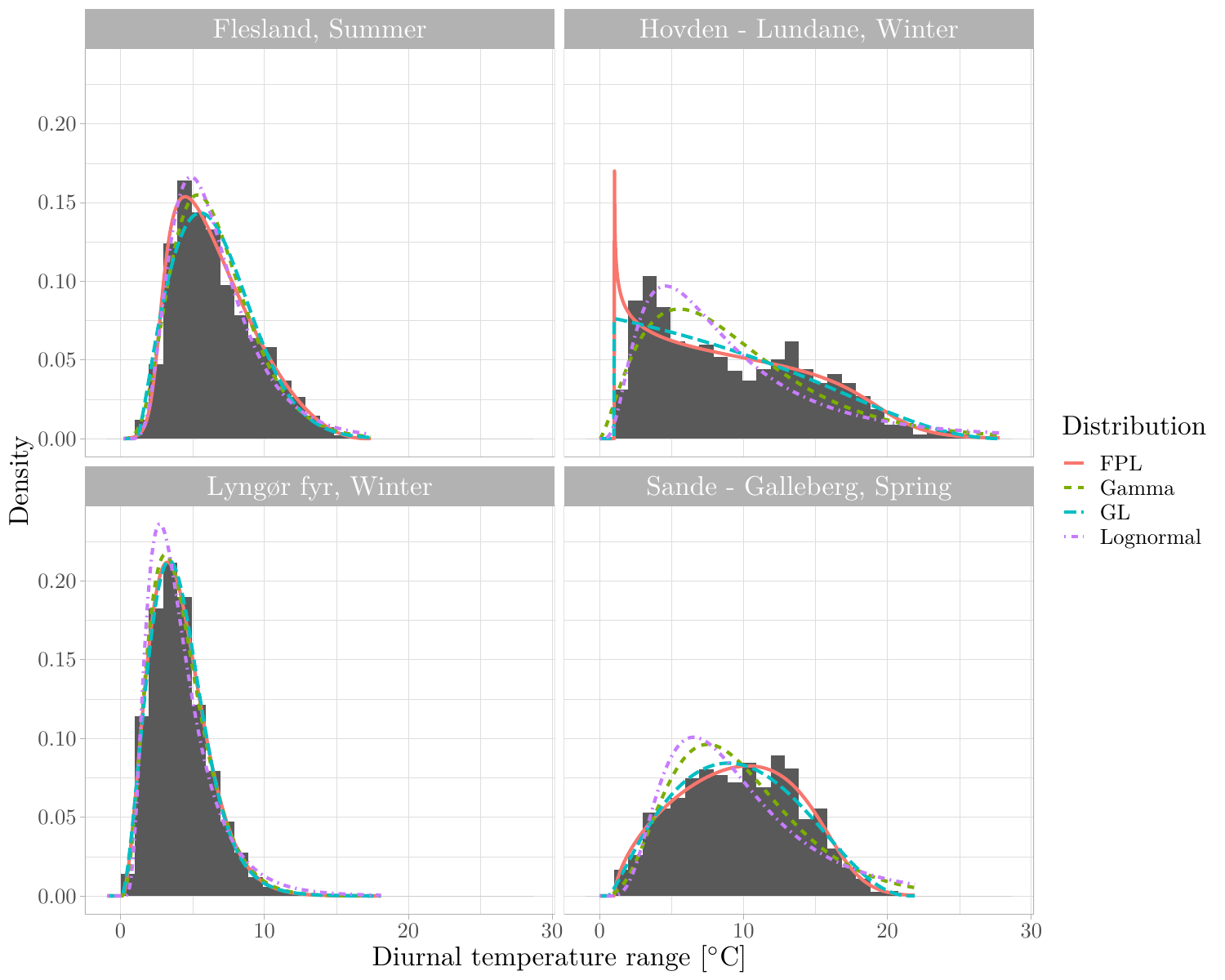}}
  \caption{Estimated probability density functions using four different parametric distribution for
    modelling diurnal temperature range. The observations of diurnal temperature range are displayed
    using histograms.}
  \label{fig:univariate-comparisons}
\end{figure}

In order to evaluate the model performance of the FPL distribution for diurnal temperature range, the
distribution is used for modelling data from southern Norway.
Diurnal temperature range is modelled separately for all seasons and weather
stations, using the FPL distribution and three other competing parametric distributions: the gamma distribution,
lognormal distribution and the generalised lambda (GL) distribution.
The gamma- and lognormal distributions are
fitted to data using maximum likelihood estimation, implemented in the \texttt{R} package
\texttt{MASS} \citep{venables02_moder_applied_statis_s}. The GL distribution
is a specialisation of the FPL distribution, parametrised with four parameters, and it is fitted using maximum likelihood, implemented in the
\texttt{gld} package. The FPL distribution is fitted to diurnal temperature range using all three inference
methods described in Section~\ref{sec:inference}.
In-sample model comparison is performed using the CRPS. This is computed numerically for the GL distribution,
using~\eqref{eq:crps}. For the gamma- and lognormal
distributions, CRPS is computed using the \texttt{scoringRules} package
\citep{jordan19_evaluat_probab_forec_with}.

Table~\ref{tab:local-crps} displays the mean CRPS over all 112 weather stations for each season
and all our chosen models.
Apart from the FPL results during summer, all three model fits with the
FPL distribution attain a lower mean CRPS than the competing models. The differences in CRPS might seem small,
but a simple permutation test shows that there is a statistically significant difference between the scores of the
FPL and GL distributions, except during summer when inference is performed with the method of quantiles. The same permutation test finds no
evidence that there is a difference in CRPS when using the method of quantiles and the starship
method, but there is some evidence that both the starship and the method of quantiles attains
better model fits than the maximum likelihood estimation.
The mean computation time for estimating the FPL parameters at a single
location is approximately 0.3 seconds with the method of quantiles, 2.7 seconds with maximum
likelihood estimation and 4.0 seconds with the starship method, meaning that it takes two
minutes to estimate parameters for all stations and seasons with the method of quantiles, and
thirty minutes with the starship method.
\begin{table}
  \centering
  \renewcommand*{\arraystretch}{1.5}
  \caption{Mean CRPS over all 112 weather stations during each season. Four different distributions
    are fitted to diurnal temperature range data. The FPL distribution are fitted to data using the method of
    quantiles (MQ), maximum likelihood (ML) estimation, and the starship method. The best mean CRPS
    for each season is written in \textbf{bold}.} 
  \label{tab:local-crps}
  \begin{tabular}{lllllll}
    \toprule 
    Season & FPL (MQ) & FPL (ML) & FPL (starship) & GL & Gamma & Lognormal \\
    \midrule 
    Winter & \(\bm{1.577}\) & \(1.577\) & \(1.577\) & \(1.578\) & \(1.580\) & \(1.589\) \\
    Spring & \(2.109\) & \(2.109\) & \(\bm{2.108}\) & \(2.111\) & \(2.122\) & \(2.141\) \\
    Summer & \(1.851\) & \(1.849\) & \(\bm{1.849}\) & \(1.851\) & \(1.857\) & \(1.867\) \\
    Autumn & \(\bm{1.715}\) & \(1.715\) & \(1.715\) & \(1.717\) & \(1.720\) & \(1.732\) \\
    \bottomrule 
  \end{tabular}
\end{table}
Figure~\ref{fig:univariate-comparisons} displays the
fitted probability density functions
of all models at the four stations from Figure~\ref{fig:range-histograms}. The FPL distribution is fitted to
data using the method of quantiles. It seems that the flexibility of the
FPL distribution makes it able to model the many shapes of diurnal temperature range better than the competing
models. Especially in the lower right plot one can see how the added flexibility of the FPL distribution allows
it to provide a slightly better fit to data than the GL distribution, which clearly is the strongest competitor.
The FPL distribution attains a lower CRPS than the competing models in all four sub-plots.

\begin{figure}
  \centerline{\includegraphics[width=.8\linewidth]{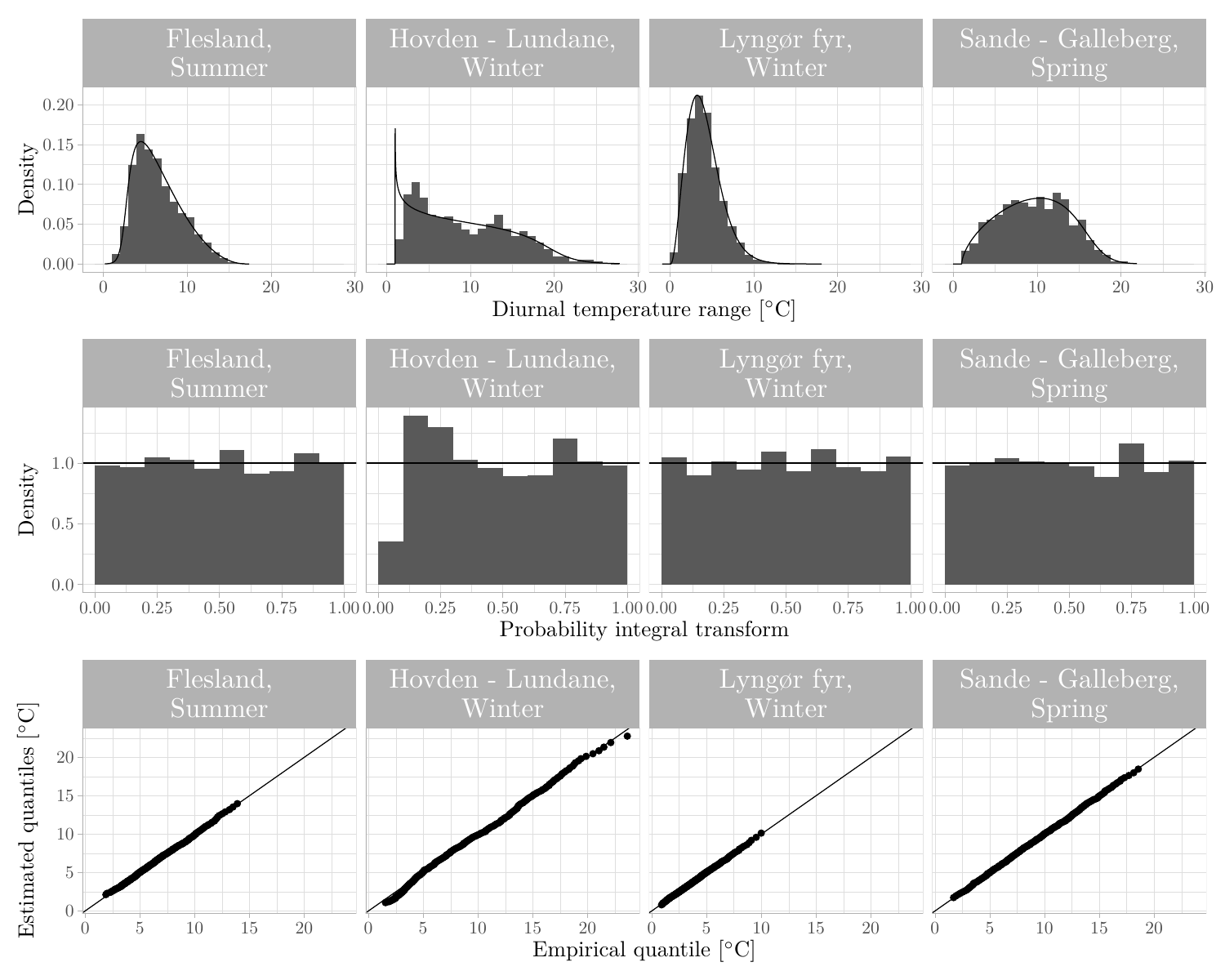}}
  \caption{
    Upper plots: Histograms displaying observed diurnal temperature
    range are plotted along with the probability density functions of the
    FPL distribution.
    Middle plots: Histograms displaying the PIT of diurnal temperature range with the estimated FPL parameters.
    Lower plots:
    QQ-plots displaying sample quantiles of diurnal temperature range
    against distributional quantiles of the estimated FPL distributions.
  }
  \label{fig:local_distribution_fits}
\end{figure}

The FPL model fits are further evaluated to assess absolute
performance. For the remainder of the paper we choose to use the method of quantiles for fitting
the FPL distribution to diurnal temperature range data, as all three inference methods perform almost equally
well, and the method of quantiles is necessary for the regression model described in
Section~\ref{sec:quantile_regression}. Figure~\ref{fig:local_distribution_fits} displays different
properties of the model fit of the FPL distribution at the four 
stations from Figure~\ref{fig:univariate-comparisons}.
It is evident that the FPL distribution provides a good model fit to observed diurnal temperature range data.
The model struggles slightly with the bimodal
distribution at Hovden - Lundane, but has an excellent fit to the data at the other three
stations. A visual assessment of the model fit for all other stations finds that the results in
Figure~\ref{fig:local_distribution_fits} are representative for most of the available data.
The mean PIT errors \(e_\mu\) and \(e_\sigma\) are displayed in Table~\ref{tab:crps-table}. Both
are of magnitude \(10^{-4}\) for the marginal FPL model, implying high overall performance.

\begin{table}
  \centering
  \renewcommand*{\arraystretch}{1.5}
  \caption{Mean CRPS and PIT errors \(e_\mu\) and \(e_\sigma\) over all 112 weather stations,
    computed for the marginal model and the regression model. The regression model
    is fitted to data both in-sample, and out-of-sample using leave-one-out cross-validation.}
  \label{tab:crps-table}
  \begin{tabular}{llcrr}
\toprule 
Season & Model & CRPS & \multicolumn{1}{c}{\(e_\mu\)} & \multicolumn{1}{c}{\(e_\sigma\)} \\
\midrule 
Winter & Marginal & \(1.58\) & \( 0.00 \cdot 10^{-2}\) & \( 0.00 \cdot 10^{-2}\) \\
 & Regression, in-sample & \(1.63\) & \( 1.46 \cdot 10^{-2}\) & \(-0.30 \cdot 10^{-2}\) \\
 & Regression, out-of-sample & \(1.63\) & \( 1.44 \cdot 10^{-2}\) & \(-0.35 \cdot 10^{-2}\) \\
\midrule 
Spring & Marginal & \(2.11\) & \( 0.06 \cdot 10^{-2}\) & \(-0.05 \cdot 10^{-2}\) \\
 & Regression, in-sample & \(2.29\) & \( 1.10 \cdot 10^{-2}\) & \(-1.35 \cdot 10^{-2}\) \\
 & Regression, out-of-sample & \(2.31\) & \( 1.08 \cdot 10^{-2}\) & \(-1.49 \cdot 10^{-2}\) \\
\midrule 
Summer & Marginal & \(1.85\) & \( 0.01 \cdot 10^{-2}\) & \( 0.01 \cdot 10^{-2}\) \\
 & Regression, in-sample & \(2.12\) & \( 1.70 \cdot 10^{-2}\) & \(-3.51 \cdot 10^{-2}\) \\
 & Regression, out-of-sample & \(2.16\) & \( 1.58 \cdot 10^{-2}\) & \(-3.74 \cdot 10^{-2}\) \\
\midrule 
Autumn & Marginal & \(1.71\) & \( 0.03 \cdot 10^{-2}\) & \(-0.05 \cdot 10^{-2}\) \\
 & Regression, in-sample & \(1.76\) & \( 0.97 \cdot 10^{-2}\) & \(-0.54 \cdot 10^{-2}\) \\
 & Regression, out-of-sample & \(1.77\) & \( 0.94 \cdot 10^{-2}\) & \(-0.60 \cdot 10^{-2}\) \\
\bottomrule 
  \end{tabular}
\end{table}

An overview of the estimated FPL parameters is given in Table~\ref{tab:param-table}. The location
and scale parameters are largest during summer and spring, and the tail parameters
\(\lambda_4\) and \(\lambda_5\) seem to take approximately the same values for all
seasons. The estimator for \(\lambda_5\) is mostly far away from \(-0.5\), indicating that the
restriction of \(\lambda_5 > -0.5\) has not lead to a decrease in model performance. The tail
weight \(\lambda_3\) seems to be almost evenly distributed between \(-1\) and \(1\), but its
distribution is clearly most focused on the positive side, where it lends most weight to the right
tail of the FPL distribution.
When examining the marginal parameter estimates in a map (results not shown) we find that both
\(\hat{\lambda}_1^*\) and \(\hat{\lambda}_4\) increase when moving eastwards. The opposite is
found for \(\hat{\lambda}_3\), which attains its largest values to the west. \(\hat{\lambda}_2^*\) and
\(\hat{\lambda}_5\) take on low values along the coast, and increase as we move further away from
the sea, and further to the east. There are some locations where the estimates for \(\lambda_4\) and \(\lambda_5\) are much
larger than 1. This is most likely caused by the problems discussed in
Section~\ref{sec:simulation-study}, where a large change in a tail parameter combined with a
change in \(\lambda_3\) results in little change in the overall shape of the FPL distribution. The estimators
for \(\lambda_2^*\) and \(\lambda_3\) are also showing unusual values at these locations.

\begin{table}
  \centering
  \renewcommand*{\arraystretch}{1.5}
  \caption{The FPL parameters are estimated at all weather stations, using the marginal model and
    the out-of-sample regression model. Median parameter estimates over all 112 locations are displayed,
    along with the \(2.5\%\) and the \(97.5\%\) quantiles.}
  \label{tab:param-table}
  \resizebox{\textwidth}{!}{%
  \begin{tabular}{llrrlrrlrrlrrl}
    \toprule 
    & & \multicolumn{3}{c}{Winter} & \multicolumn{3}{c}{Spring} & \multicolumn{3}{c}{Summer} & \multicolumn{3}{c}{Autumn} \\
    \cmidrule(lr){3-5} \cmidrule(lr){6-8} \cmidrule(lr){9-11} \cmidrule(lr){12-14} 
    Model & \(\hat{\bm{\lambda}}\) & \(2.5\%\) & \(50.0\%\) & \(97.5\%\) & \(2.5\%\) & \(50.0\%\) & \(97.5\%\) & \(2.5\%\) & \(50.0\%\) & \(97.5\%\) & \(2.5\%\) & \(50.0\%\) & \(97.5\%\) \\
    \midrule 
    Marginal & \(\hat{\lambda}_1\) & \( 3.0\) & \(5.0\) & \( 7.7\) & \( 3.6\) & \(8.0\) & \(11.5\) & \( 3.7\) & \(8.8\) & \(11.8\) & \( 3.1\) & \(5.6\) & \( 7.4\) \\
    & \(\hat{\lambda}_2\) & \( 1.9\) & \(3.7\) & \( 7.0\) & \( 2.3\) & \(5.7\) & \( 8.1\) & \( 1.9\) & \(4.9\) & \( 7.1\) & \( 1.9\) & \(4.4\) & \( 6.9\) \\
    & \(\hat{\lambda}_3\) & \(-0.6\) & \(0.2\) & \( 1.0\) & \(-0.9\) & \(0.1\) & \( 1.0\) & \(-0.6\) & \(0.4\) & \( 1.0\) & \(-0.7\) & \(0.6\) & \( 1.0\) \\
    & \(\hat{\lambda}_4\) & \( 0.0\) & \(0.4\) & \( 0.8\) & \( 0.0\) & \(0.4\) & \( 1.2\) & \( 0.0\) & \(0.2\) & \( 0.5\) & \( 0.0\) & \(0.2\) & \( 1.1\) \\
    & \(\hat{\lambda}_5\) & \(-0.1\) & \(0.1\) & \( 0.3\) & \(-0.2\) & \(0.2\) & \( 0.5\) & \(-0.1\) & \(0.3\) & \( 0.6\) & \( 0.0\) & \(0.2\) & \( 0.5\) \\
    \midrule 
    Regression & \(\hat{\lambda}_1\) & \( 2.9\) & \(4.8\) & \( 7.1\) & \( 3.4\) & \(7.8\) & \(12.5\) & \( 4.6\) & \(8.3\) & \(11.2\) & \( 3.3\) & \(5.5\) & \( 7.6\) \\
    & \(\hat{\lambda}_2\) & \( 1.5\) & \(3.8\) & \( 6.2\) & \( 2.3\) & \(5.7\) & \( 8.7\) & \( 3.6\) & \(5.2\) & \( 7.1\) & \( 1.9\) & \(4.6\) & \( 6.7\) \\
    & \(\hat{\lambda}_3\) & \(-0.5\) & \(0.2\) & \( 0.9\) & \(-0.5\) & \(0.5\) & \( 1.0\) & \( 0.0\) & \(0.6\) & \( 0.9\) & \(-0.3\) & \(0.5\) & \( 0.8\) \\
    & \(\hat{\lambda}_4\) & \( 0.0\) & \(0.4\) & \( 0.6\) & \( 0.0\) & \(0.3\) & \( 0.7\) & \( 0.0\) & \(0.1\) & \( 0.5\) & \( 0.0\) & \(0.4\) & \( 0.8\) \\
    & \(\hat{\lambda}_5\) & \(-0.3\) & \(0.0\) & \( 0.3\) & \(-0.3\) & \(0.2\) & \( 0.7\) & \( 0.0\) & \(0.3\) & \( 0.7\) & \(-0.2\) & \(0.2\) & \( 0.4\) \\
    \bottomrule 
  \end{tabular}
}
\end{table}

\subsection{Regression model}
\label{sec:qr_results}

\begin{figure}
  \centerline{\includegraphics[width = .9\linewidth]{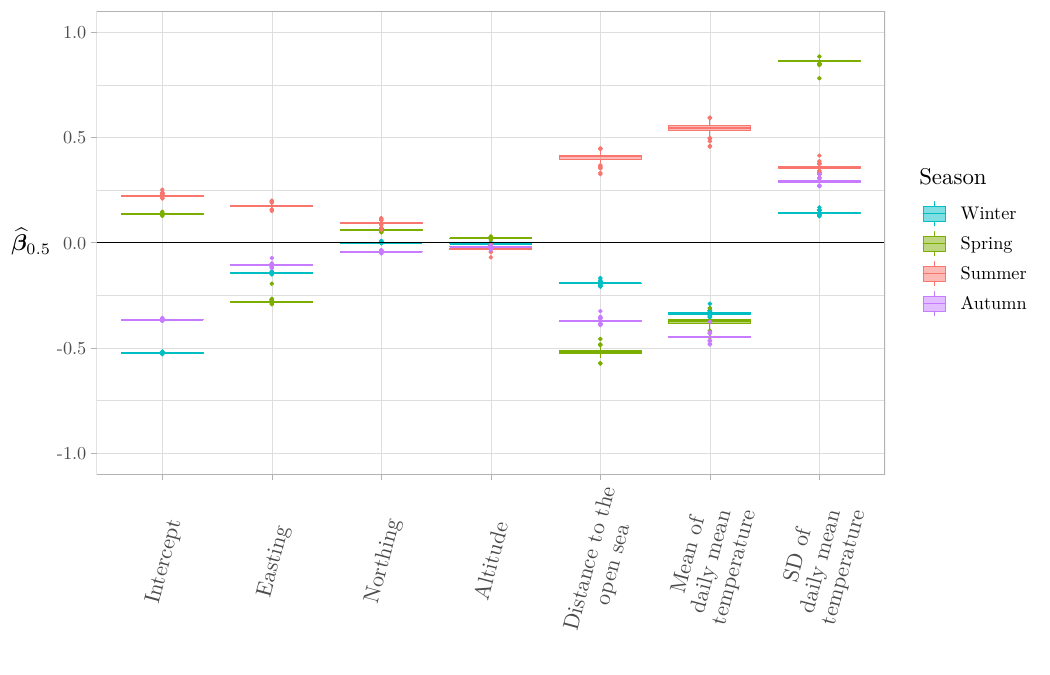}}
  \caption{Estimated regression coefficients
    \(\widehat{\bm{\beta}}_{0.5}\) for the median regression. The
    parameter estimation is performed out-of-sample, resulting in 112
    parameter estimates within each season. The whiskers in the box-plot have a maximum length
    of \(1.5\) times the interquartile range. The \(y\)-axis has unit ``standard deviations of
    \(\bm{y}\)'', meaning that if a regression coefficient is e.g.\ 0.4, then a change of one
    standard deviation for the corresponding explanatory variable causes a change of \(0.4 \cdot
    \text{SD}(\bm{y})\) in the median of \(\bm{y}\).}
  \label{fig:qr_boxes}
\end{figure}

The regression model is applied for modelling diurnal temperature range for each season
  separately, using all the explanatory variables introduced in Section~\ref{sec:data}. Before we
  apply the regression model, each
  explanatory covariate is standardised to have zero mean and a standard deviation of one. Modelling
is performed in-sample using all available data, and out-of-sample in a leave-one-out
cross-validation study. Thus, in the cross-validation study, the regression coefficients
\(\bm{\beta}_{p_i},\ i = 1, 2, \ldots, 99,\)
are estimated 112 times for each season, each time by leaving one station out of the training data. The FPL
parameters are then estimated at the one station that was not included in training the quantile
regression models. Table~\ref{tab:crps-table} shows little difference in performance between
in-sample and out-of-sample estimation, indicating that our model does not overfit to the
data. For winter and autumn data, the differences between the CRPS of the marginal
model and the regression model are small. However, during summer and spring, there is a considerable
difference in performance between the two models. The calibration of the regression model is clearly
worse than that of the marginal model for all seasons. However, PIT errors with a magnitude of
\(10^{-2}\) is still good, even though it is worse than a magnitude of \(10^{-4}\).
All estimated regression coefficients for the 112 out-of-sample median regressions are displayed in
Figure~\ref{fig:qr_boxes}. The variability in \(\widehat{\bm{\beta}}_{0.5}\) within each season is
small, indicating that the estimation procedure is robust against minor changes in the training
data. We see that the most influential explanatory variables across
all seasons are the distance to the open sea and the historical temperature
observation. There is a large difference between summer, spring and the other two
seasons for these explanatory variables. For example,
a larger mean temperature will lead to larger median range during summer, but lower median range
during all other seasons.
This might be connected to the difference in model performance during
summer and spring. Similar trends are found in the \(\widehat{\bm{\beta}}_p\) for all other
probabilities \(p\).

\begin{figure}
  \centerline{\includegraphics[width=.8\linewidth]{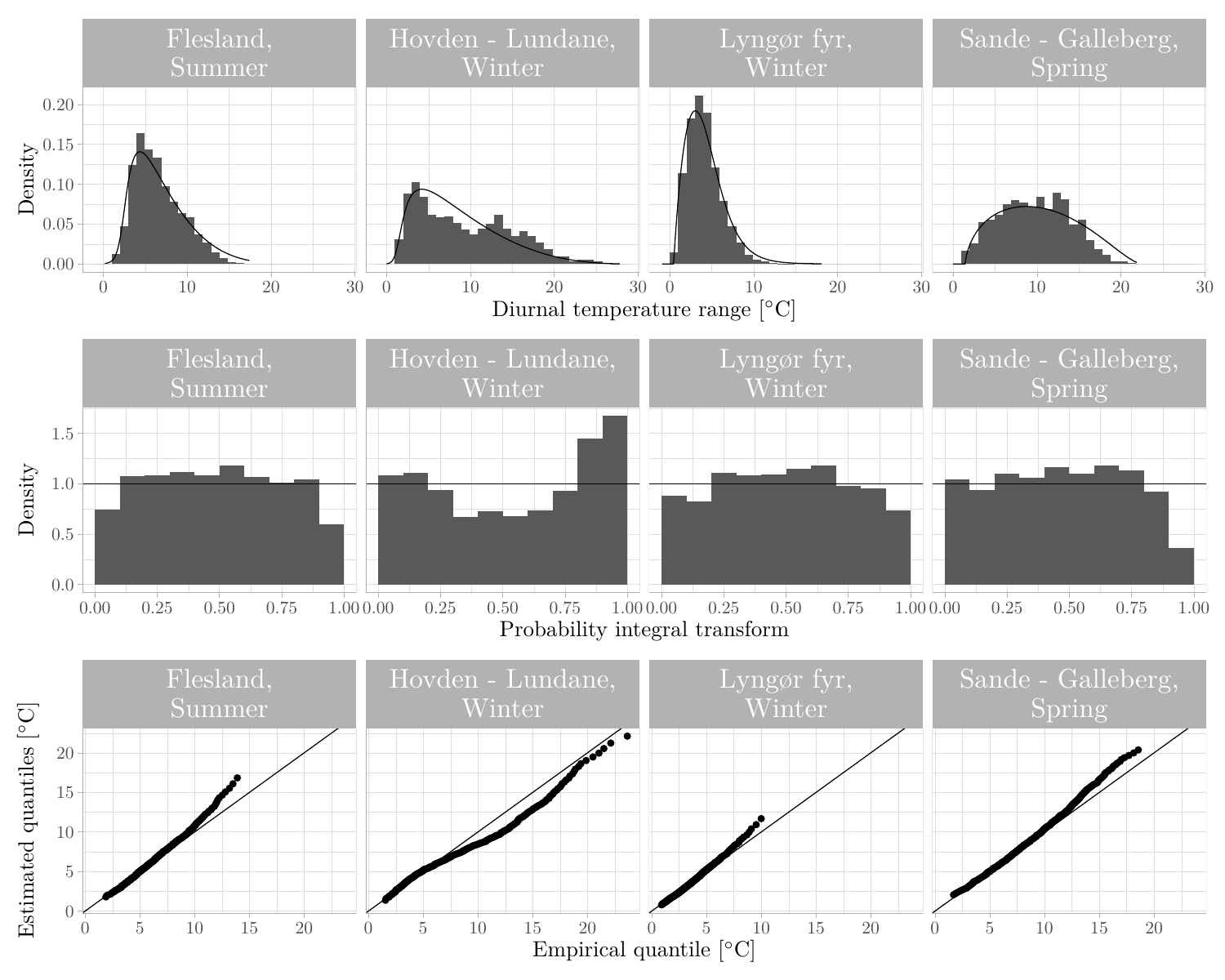}}
  \caption{
    Upper plots: Histograms displaying observed diurnal temperature
    range are plotted along with the probability density functions of the
    FPL distribution from the out-of-sample regression model.
    Middle plots: Histograms displaying the PIT of diurnal temperature range with the estimated FPL parameters.
    Lower plots:
    QQ-plots displaying sample quantiles of diurnal temperature range
    against distributional quantiles of the estimated FPL distributions.
    \label{fig:regional-out-of-sample}
  }
\end{figure}

Figure~\ref{fig:regional-out-of-sample} displays the out-of-sample
estimation results for the same stations and seasons as in
Figure~\ref{fig:local_distribution_fits}. While the regression
model is able to capture the overall shapes at each location, some
deviations are noticeable. In particular, the estimated right tail is too heavy at all stations but
Hovden - Lundane, where it is too light. Apart from this, the model fits seem adequate for the bulk of the data.
As seen in Table~\ref{tab:param-table}, the estimated FPL parameters from the out-of-sample
regression model shares many similarities with the parameter estimates from the marginal model. Most of the
spatial trends from the marginal model fits are also found when examining the estimates from the
regional model.

\begin{figure}
  \hspace{2em}
  \centerline{\includegraphics[width = .95\linewidth]{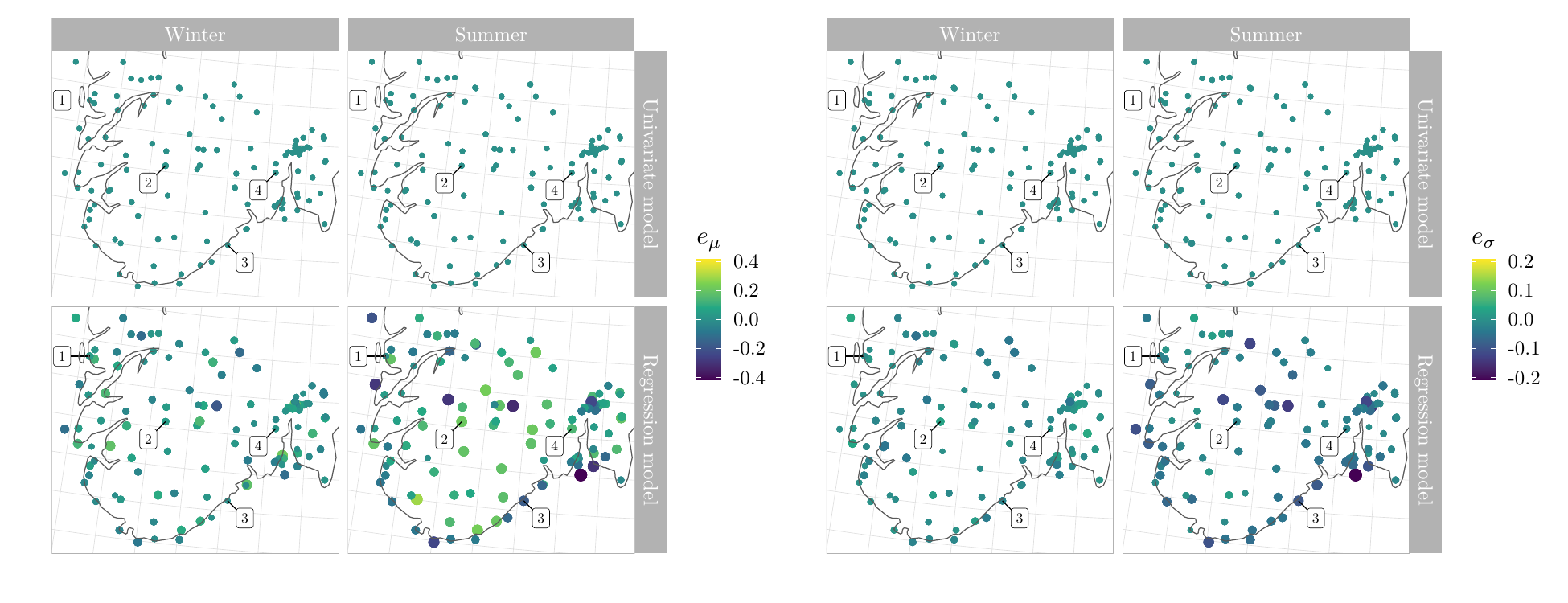}}
  \caption{PIT mean errors \(e_\mu\) and standard deviation errors
    \(e_\sigma\) are displayed for the regression model and the marginal
    model, during summer and winter. The magnitude of each error is
    represented by the radius of each dot. Values of each error are
    represented by the colour of each dot.
    The locations of the four stations from
    Figures~\ref{fig:range-histograms}, \ref{fig:local_distribution_fits} and
    \ref{fig:regional-out-of-sample} are presented in all the plots. These
    are: 1) Flesland, 2) Hovden - Lundane, 3) Lyngør fyr and 4) Sande - Galleberg.
  }
  \label{fig:pit_plots}
\end{figure}

Figure~\ref{fig:pit_plots} summarises the calibration of both the
marginal model fits and the out-of-sample regression model fits for
summer and winter seasons by displaying the PIT errors \(e_\mu\)
and \(e_\sigma\) at all locations. For the out-of-sample assessment of the regression model, the
calibration of the estimated distributions varies substantially
between the two seasons. In winter, the calibration is,
expectedly, somewhat worse than that of the marginal model. However,
only a few stations show considerable lack of calibration. Both
positive and negative values of \(e_\mu\) are observed, indicating both
negative and positive biases. However, the values of \(e_\sigma\) are
rather negative than positive, indicating a slight tendency
towards overdispersion or too large spread. Similar patterns are
observed for autumn (results not shown). The performance of the regression model in summer is considerably
worse than that in winter. There is a distinct jump in the values of \(e_\mu\) and \(e_\sigma\) as the
distance from the sea increases. The jump in \(e_\mu\) implies that we observe mostly positive
biases along the coast and negative biases at inland locations.
Figure~\ref{fig:median_values} shows
that there is a similar pattern of change in the median of diurnal
temperature range along the coast and further inland during summer and spring. This indicates
that the regression model is too smooth, and fails to model the
transition from coastal to inland climate. A similar
jump in \(e_\mu\) and \(e_\sigma\) can be seen for spring data,
although the difference is not as considerable as for the summer
data.

Figure~\ref{fig:covariates} hints that there is more
information to gain from the climatological explanatory variables
describing the distribution of mean temperature in winter and autumn
than in summer and spring. In order to investigate whether the
seasonal difference in skill is due to this effect, we repeated the
regression analysis using only the four geographic and orographic
explanatory variables. Under this model, the performance for winter and autumn data is considerably
worse, while it stays almost unchanged for spring and summer data (results not shown). The magnitude
of the PIT errors
during winter and autumn with this model is also comparable with those for spring and summer data in
the original model.

\section{Conclusion and discussion}
\label{sec:discussion}

This paper proposes to use the five-parameter lambda (FPL) distribution
to model the distribution of diurnal temperature range. A distributional quantile regression model is also
proposed, where diurnal temperature range is modelled using a combination of quantile regression and
marginal modelling with the FPL distribution.
Parameter estimation is performed using the method of quantiles, which is a fast inference
method that compares well with the competing
inference methods of maximum likelihood estimation and the starship method.
Diurnal temperature range from southern Norway is modelled using the marginal FPL model and the regression
model. The marginal FPL model provides a good fit to diurnal temperature range data,
and the regression model shows promise and is able to capture much, but not all, of the spatial
trends in the distribution of diurnal temperature range.

We propose a reparametrisation of the FPL distribution
that allows for easier inference by directly connecting the location \(\lambda_1\) and scale \(\lambda_2\) to quantiles
of the distribution. This allows us to select better initial values for parameter estimation of the
FPL parameters. The reparametrisation also limits the parameter space of $\bm{\lambda}$ so that
$\lambda_5 > -1/2$. This does not appear to influence our results.

Although the method of quantiles shows great success in fitting the FPL distribution to data, we experience
situations where the estimators for \(\lambda_4\) and \(\lambda_5\) become too large with
respect to the true values, which the method accounts for by providing biased estimates for
\(\lambda_3\). For large sample sizes this does not seem to affect
the model fit much, but for smaller sample sizes it seems to negatively affect the performance.
Further work should be put into understanding the effect of the parameters of the FPL distribution, and why the method
of quantiles overestimates tail parameters in certain situations.

Model evaluation is mainly performed using the probability integral transform (PIT) and the
continuous ranked probability score (CRPS). A novel closed-form expression for the CRPS with an
FPL forecast distribution is developed, which makes model evaluation faster and simpler. Other scoring rules may
be more appropriate for evaluating model fit in
larger and more inhomogeneous regions \citep{bolin19_scale_depen}, but we believe the CRPS to be a
good choice for model evaluation in the current setting.

While the regression model is not able to fully capture the spatial
patterns in diurnal temperature range for spring and summer, its
performance is promising, especially for winter and autumn data. It is our belief that one
can achieve better results with an improved selection of explanatory
variables and, potentially, further development of the distributional quantile
regression method. As an example,
many of the weather stations with high PIT errors \(e_\mu\) and
\(e_\sigma\) are located 
along the coast. Consequently, it might be reasonable
to better distinguish between coastal observations and inland observations,
e.g. through the introduction of a binary explanatory variable. A 
transformation of variables could also improve our models. The relationships
between the median of diurnal temperature range and the available
explanatory variables shown in Figure~\ref{fig:covariates} do not seem linear for
the geographical information. One might find possible transformations of
the explanatory variables which are able to improve the linear relationships
between the quantiles of diurnal temperature range and the available
explanatory variables.
In addition to the explanatory variables used in the current study,
one might find important dependencies
between diurnal temperature range and other climate variables, such as
daily precipitation, wind speed and the degree of cloud cover. Especially
precipitation and cloud cover have been found to be highly negatively
correlated with diurnal temperature range
\citep{zhou2009spatial,waqas2018observed}. It is not obvious how such
explanatory variables should be incorporated in our model, though.
The regression model for diurnal temperature range
only includes spatial fixed effects. The inclusion of spatial random
effects might improve model performance
\citep{lum2012spatial,self21_ident_meteor_driver_pm2}. Additionally, attempts to estimate parameter
uncertainty will be affected by the temporal autocorrelation of diurnal temperature range, which
is statistically significant for lags up to approximately one week.
Consequently, further modelling attempts should also 
aspire to include a temporal framework.
A comprehensive modelling framework for temperature range should furthermore include modelling
components to account for data issues such as data inhomogeneities and measurement errors, including
recording precision. This is
particularly important in settings where long data series from various
data sources are combined into a single analysis, like in the unified Bayesian approach that was
introduced in the EUSTACE project \citep{EUSTACE2020}. In this project temporal and spatial autocorrelations
were handled by latent random field components, and the data inhomogeneities were directly estimated via independent random effect
variables. That approach allowed the propagation of uncertainty through the entire analysis
system, in contrast to more traditional data homogenisation methods that handle this as a separate
pre-processing step. In principle, the FPL model for diurnal temperature range can
be incorporated into the observation level of such hierarchical models.

Two of the explanatory variables in the regression model are based on
historical daily mean temperature. For spatial interpolation one might
argue that observations of daily mean temperature are unavailable at most
locations where there are no observations of diurnal temperature range.
However, while the literature on modelling diurnal
temperature range is lacking, much effort and success has been put
into the modelling of mean temperature \citep[e.g.][]{maraun2018,
  haylock2008eobs}. We assume that there already exist satisfactory
spatial and temporal models for Norway, which are able to describe the
historical mean and variance of daily mean temperature between 1989
and 2018 with a high performance. The Nordic Gridded Climate Data Set
version 2 \citep{lussana2018temp}, e.g., models daily mean
temperature with high performance everywhere in Norway, Finland and Sweden. Accordingly, all
explanatory variables can be provided at any location in Norway. 

Better models for diurnal temperature range may be important for improving interpolation and
statistical downscaling of temperature projections from climate models
\citep[e.g.][]{maraun2018}.
The common approach today is to perform separate modelling of daily
maximum, minimum and mean temperature. However, this can lead to inconsistencies such as predictions
where the daily mean is larger than the daily maximum temperature
\citep[e.g.][]{lussana19_daily_precip_temper_datas_over_norway}. In addition, the three temperature
variables are heavily dependent and should be modelled jointly, but multivariate modelling is often too
challenging and computationally demanding. An alternative method for modelling these three
temperature variables is to transform minimum, maximum and mean temperature to diurnal temperature
range, mean temperature, and the location of the daily mean inside the temperature range. This would
remove all ordering inconsistencies between minimum, mean and maximum temperature, and it would
considerably reduce correlations between the three variables. Analysis of the data used in this
paper finds that the absolute values of pairwise sample correlations between daily minimum, maximum
and mean temperature mainly lie in the interval $[0.9,1]$. Sample correlations
between diurnal temperature range, daily mean temperature and the location of the mean inside the
range, on the other hand, are always below $0.5$. Thus, appropriate statistical models for diurnal
temperature range and its relationship with daily mean temperature can be of great interest as model
components for multivariate temperature modelling approaches. Further work should be conducted to
examine this approach.

\section*{Acknowledgments}
We thank Sara Martino, Cristian Lussana, Irene Brox Nilsen and Erik Kjellstr\"om for
helpful discussions. Thordis L. Thorarinsdottir acknowledges the
support of the Research Council of Norway through project nr. 255517
``Post-processing Climate Projection Output for Key Users in
Norway''.
%% Mandated EUSTACE funding statement:
As part of the EUSTACE project, Finn Lindgren received funding
from the European Union's ``Horizon 2020 Programme for Research and Innovation'',
under Grant Agreement no 640171.
The marginal FPL model for diurnal temperature range was first developed within the EUSTACE research project for historical
global daily temperature reconstruction \citep{EUSTACE2020}, but was not used in their final analysis
method due to constraints in the time frame of the project.
The code and data used in this paper are available at \url{https://github.com/siliusmv/FPLD}.
This article is an extension of the Master's thesis
of Silius M. Vandeskog \citep{vandeskog2019master}.

\subsection*{Conflict of interest}

The authors declare no potential conflict of interests.

\printbibliography

\end{document}